\documentclass{article}

\usepackage{PRIMEarxiv}

\usepackage[utf8]{inputenc} 
\usepackage[T1]{fontenc}
\usepackage{url}   
\usepackage{bm}
\usepackage{multicol}
\usepackage{sectsty}
\usepackage{hyperref}
\usepackage{nccmath}
\usepackage{float}
\usepackage{commath}
\usepackage{amsmath}
\usepackage{amsfonts}
\usepackage{amssymb}
\usepackage[numbers]{natbib}
\usepackage{dcolumn}
\usepackage{mathtools}
\usepackage{float}
\usepackage{fancyhdr}
\usepackage{fnpos}
\usepackage[english]{babel}
\usepackage{lastpage}
\usepackage{booktabs}
\usepackage{amsfonts}
\usepackage{mathptmx}
\usepackage{mathtools}
\usepackage{amsfonts}
\makeatletter
\def\blfootnote{\xdef\@thefnmark{}\@footnotetext}
\makeatother
\usepackage{nicefrac}
\usepackage{microtype}
\usepackage{lipsum}
\usepackage{fancyhdr}
\usepackage{graphicx} 
\graphicspath{{media/}}

\title{Numerical validation of Ehrenfest theorem in a Bohmian perspective for non-conservative systems

}

\author{Matheus M. A. Paixão\thanks{Email:\href{mailto:matheuspaixao@cbpf.br}{matheuspaixao@cbpf.br}}\hspace{0.5mm} and Henrique Santos Lima \thanks{Email:\href{mailto:hslima94@cbpf.br}{hslima94@cbpf.br}}\\ \\ \textit{Centro Brasileiro de Pesquisas F\'{\i}sicas, Rua Doutor Xavier Sigaud 150, 22290-180, Rio de Janeiro, Brazil}}

\begin{document}
\maketitle

\begin{abstract}
In this work we make a high precision numerical study of the Ehrenfest theorem using the Bohmian approach, where we obtain classical solutions from the quantum trajectories performing the Bohmian averages. We analyse the one-dimensional  quantum harmonic  and Duffing oscillator cases, finding numerical solutions of the time-dependent Schrödinger equation and the guidance equation for different sets of initial conditions and connects these results with the corresponding classical solutions. We also investigate the effect of introducing external forces of three types: a simple constant force, a fast-acting Gaussian impulse, and an oscillatory force with different frequencies. In the last case the resonance in the quantum trajectories was observed.
\end{abstract}

\section{Introduction}
The de Broglie-Bohm interpretation of quantum mechanics \cite{deBroglie,Bohm1,Bohm2} has been frequently studied in the last decades, due to its wide applicability and its ability to dialogue with different areas of physics. The quantum-classical treatment of quantum systems allows us to study  many topics, for instance, quantum chaos \cite{Wis2005,Conto2006,SanjibFring2013, Ivanov2019, Conto2020,Drezet2021,TzemosContopoulos20222}, quantum synchronization \cite{Wenlinli2022}, and quantum hydrodynamics \cite{Tsekov2018,LiceaSchuch2021,Licea5,LiceaSchuch20222,Frumkin2022}. Bohmian mechanics is useful to comprehend the dynamics of molecules \cite{AvanziniGiorgio2018} , the strong-field enhanced ionization \cite{Weietal2013,HosseinTelnov2013,Yangetal2013,Sawadaetal2014,Lilu2022}, entanglement \cite{Borios2013, TzemosContopoulos2020,TzemosContopoulos20223}, and scattering processes \cite{Oleg2001,Santana2021} as well. Moreover, this alternative interpretation is useful to cosmology, since it solves the measurement problem and enable the understanding of cosmological quantum singularities \cite{nelsonuniverse2021,bounce2021}.

The aim of this work is to study the validity of Ehrenfest theorem \cite{Ehrenfest1922} through a Bohmian perspective in some  simple non-conservative systems, which is a barely addressed topic in the literature \cite{Alonso2001}. In this sense, the validation of such theorem fulfills the investigations of the equivalence between the Copenhagen and this alternative interpretation. For the one-dimensional quantum harmonic oscillator, for example, we know that classical laws emerge  when we consider the mean values of quantum operators. From a Bohmian point of view, however, such averages are computed,  in the quantum equilibrium regime~\cite{Valentini1991}, over a set of initial positions distributed according to the probability  density $|\Psi(x,t=0)|^2$. Each initial condition generates a distinct trajectory, which is the solution of the guidance equation. Since the
Bohmian mechanics is built based on a classical interpretation of the quantum particles dynamics, we expect that the average value of a considerable number of possible trajectories would follow a classical law.

For this purpose, we investigate the dependency of the trajectories and oscillation amplitudes  over the initial state $\Psi(x,0)$ which we assume to be a superposition of $n$ eigenstates of the harmonic oscillator properly normalized. 
We also verify the effect of including time-dependent external forces of three types: a simple constant force, a fast-acting Gaussian impulse, and a sinusoidal with different frequencies. 

\section{Bohmian Mechanics}
In Bohmian mechanics (or pilot wave interpretation), the trajectories of quantum systems are guided by a wave function $\Psi(\vec{x},t)=R(\vec{x},t) e^{iS(\vec{x},t)/\hslash}$ through the following relation
\begin{ceqn}
\begin{equation}
    \frac{d\vec{x}}{dt}=\frac{\Vec{\nabla} S}{m},
    \label{guidance equation}
\end{equation}
\end{ceqn}
where $R(\Vec{x},t)$ and $S(\Vec{x},t)$ are the radial part and  phase of $\Psi$, respectively.

Given a set of initial positions, we can integrate the  Eq. \eqref{guidance equation} and obtain the trajectories of the particles at any instant. Inserting the previous wave function into the Schrödinger equation we have two real expressions
\begin{ceqn}
\begin{align}
    \frac{\partial S}{\partial t}+\frac{(\Vec{\nabla}S )^2}{2m} + V + Q &=0,\label{Hamilton-Jacobi}\\
   \frac{\partial R^2}{\partial t} + \Vec{\nabla}\cdot\left( R^2\frac{\Vec{\nabla} S}{m}\right)&=0.\label{continuity}
\end{align}
\end{ceqn}
The first one can be interpreted as a Hamilton-Jacobi equation for $S(\vec{x},t)$ with a supplementary potential $Q(\vec{x},t)$, called quantum potential, given by $Q=-\frac{\hslash^2}{2m}\frac{\nabla^2R}{R}$.
The second expression is a continuity equation where $R^2$ is a probability density and $\Vec{\nabla}S/m$ is a velocity field.
\section{Ehrenfest theorem}
In traditional quantum mechanics the Ehrenfest theorem  are mathematical relations that concerns the temporal evolution of the mean values of the position and momentum operators, being similar to Hamilton's equations. For a generic operator $\hat{O}$, for instance, its average is usually defined as  $\langle \hat{O}\rangle \equiv\langle\Psi  |\hat{O}| \Psi\rangle$, where $|\Psi\rangle$ is a general state . In Bohmian mechanics, however, the average of a physical property $O$ is defined as 
\begin{align}
    \langle O(t)\rangle _{dBB}=\int|\Psi(x,t)|^2 O(x,t)dx,
\end{align}
where $|\Psi|^2=R^2$ is the trajectory probability density. The quantities from which we perform the averages may have an intrinsic quantum contribution, namely the quantum potential. For a significant large number of initial positions we can approximate the mean value by  $\langle O(t)\rangle _{dBB}\approx\dfrac{1}{N} \sum_{i=1}^NO_i(t)$, with $N$ the total number of trajectories \cite{Holland1993,Wuetal2013}. 

The Bohmian one-dimensional  version of the Ehrenfest theorem ~\cite{Holland1993, Salvatore2013} is given by 
\begin{align}
    m\frac{d}{dt}{\langle x \rangle}_{dBB}&=\langle p \rangle_{dBB},\label{Ehrenfest dBB1}\\
    \frac{d}{dt}{\langle p \rangle}_{dBB}&=-\left\langle \frac{d V}{dx} \right\rangle_{dBB},
    \label{E}
\end{align}
where we substitute the operators brackets of the usual version by Bohmian averages. Combining Eqs. \eqref{Ehrenfest dBB1} and \eqref{E} we find the Newton's second law  for the quantum harmonic oscillator, since, for this case, $\langle V(x) \rangle_{dBB}=V\left(\langle x \rangle_{dBB}\right)$. It is predicted  that the  addition of an external force that only depends on time does not change this result. In fact, we simply need to add this force to the classic potential contribution, as we usually do in classical mechanics. In the next sections we will study such effect introducing  different types of force.

\section{Model and methods}
Let us consider a forced  harmonic oscillator with Hamiltonian $\hat{\mathcal{H}}=\frac{\hat{p}^2}{2m}+\frac{1}{2}m \omega_0^2 \hat{x} ^2-F(t)\hat{x}$, with $\hat{x} $ and $\hat{p}$ the position and momentum operators, respectively,  and $F(t)$ being a time-dependent force. In order to make a numerical analysis, we replace temporal and spatial variables by dimensionless ones, $t\rightarrow\omega_0 t$ and $x\rightarrow x/\sqrt{\hslash/m\omega_0}$, implying that $F\rightarrow F/\sqrt{\hslash m \omega_0^3}$. 
Let us apprise that Eqs. \eqref{Hamilton-Jacobi} and \eqref{continuity} in terms of the radial part $R(x,t)$ and the phase $S(x,t)$ are, in general, hard to solve numerically due to the non-linearity of $(\nabla S)^2$ and the high number of derivatives in Eq. \eqref{Hamilton-Jacobi}. An alternative to achieve higher precision with less computational effort is obtained separating the general wave function as $\Psi(x,t)=\Phi_r(x,t)+i\Phi_i(x,t)$, where $\Phi_r$ and $\Phi_i$ are the real and imaginary part, respectively. Thus, the time-dependent Schrödinger equation $\hat{\mathcal{H}}\Psi=i\partial_t\Psi$ leads us to two linear coupled dimensionless partial differential  equations:
\begin{align}
        &-\frac{1}{2}\frac{\partial^2 \Phi_r}{\partial x^2}+\left(\frac{1}{2}x^2-F(t)x\right)\Phi_r=-\frac{\partial \Phi_i}{\partial t},\\
        &-\frac{1}{2}\frac{\partial^2 \Phi_i}{\partial x^2}+\left(\frac{1}{2}x^2-F(t)x\right)\Phi_i=\frac{\partial \Phi_r}{\partial t}.
        \label{EDP2}
\end{align}
The original polar form is recovered from $\Phi_r$ and $\Phi_i$ via $R(x,t)=\sqrt{\Phi_r^2+\Phi_i^2}$ and $S(x,t)=\arctan\left(\frac{\Phi_i}{\Phi_r}\right)$.

We solve this system using the method of lines~\cite{SarmirChudov1963, NorRama2013, Hamdi2017}, which is a method of solving PDEs consisting in discretize all the dimensions except by one, turning the problem into a system of ODEs, where we can apply the numerical techniques available. We consider the tensor-product grid technique to make the spatial discretization~\cite{Zegeling2004}. We consider Dirichlet boundary conditions for $\Psi$, demanding that $\Psi(-L,t)=\Psi(L,t)=0$, where we take $L=10$ in the most of the cases. As initial conditions we set $\Psi(x,0)$ as a combination with the same weight of the eigenstates $\psi_{\alpha}$ of the harmonic oscillator, namely, $\Psi(x,0)=\frac{1}{\sqrt{n+1}}\sum_{\alpha=0}^{n}\psi_{\alpha}(x)$. Both methods are available in  recent versions of \textit{Mathematica} software.

Each solution passes to a second step whose task is to solve the guidance equation \eqref{guidance equation}, which is a simple  ordinary differential equation (ODE), but strongly dependent on the set of initial positions and wave packets.
The ODE  is solved  with a sample between 400 and 2000 initial positions randomly distributed according to $|\Psi(x,0)|^2$, in  which  more complex forces demand a greater number of  initial points to achieve higher precision. All trajectories are included in the average of the coordinates and we use a simple nonlinear least square method  to find the function that fits better in our solutions. We fix the time step of the ODE in $dt=0.01$.

\section{Results}

We consider in our simulations  three distinct cases:
1) A constant force $F(t) = C$.
2) An impulsive force of Gaussian type that acts as a perturbation over the system, with $F(t)=\frac{1}{\sqrt{2\pi\sigma^2}} \exp{\left(-\frac{(t-t_{\mu})^2}{2\sigma^2}\right)}$.
3) A sinusoidal force of the form $F(t)=A\cos{(\omega t)}$.

Note that the pure classical solution for the general case is given by

\begin{equation}
\label{conv}
   x(t)=x(t)_{HO}+\int_{0}^{t}d\,\tau\, F(t-\tau)\sin(\tau),
\end{equation}
where the first term  is the solution of the simple harmonic oscillator, while the second one is a convolution relating the external force. Therefore, we expect for the validation of the Ehrenfest theorem something similar to Eq. \eqref{conv}, where the positions are replaced by their averages.

\subsection{Quantum harmonic oscillator ($F(t)=0$)}

\begin{figure*}
 \centering
 \includegraphics[width=6.cm]{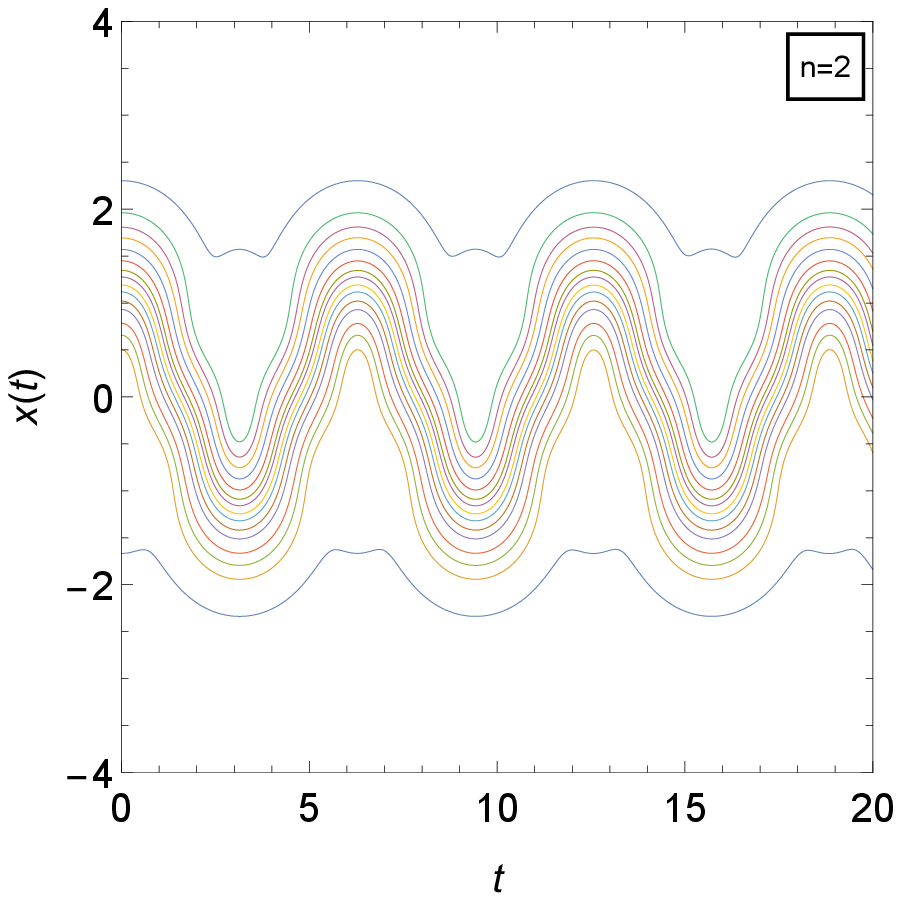}
 \includegraphics[width=6.cm]{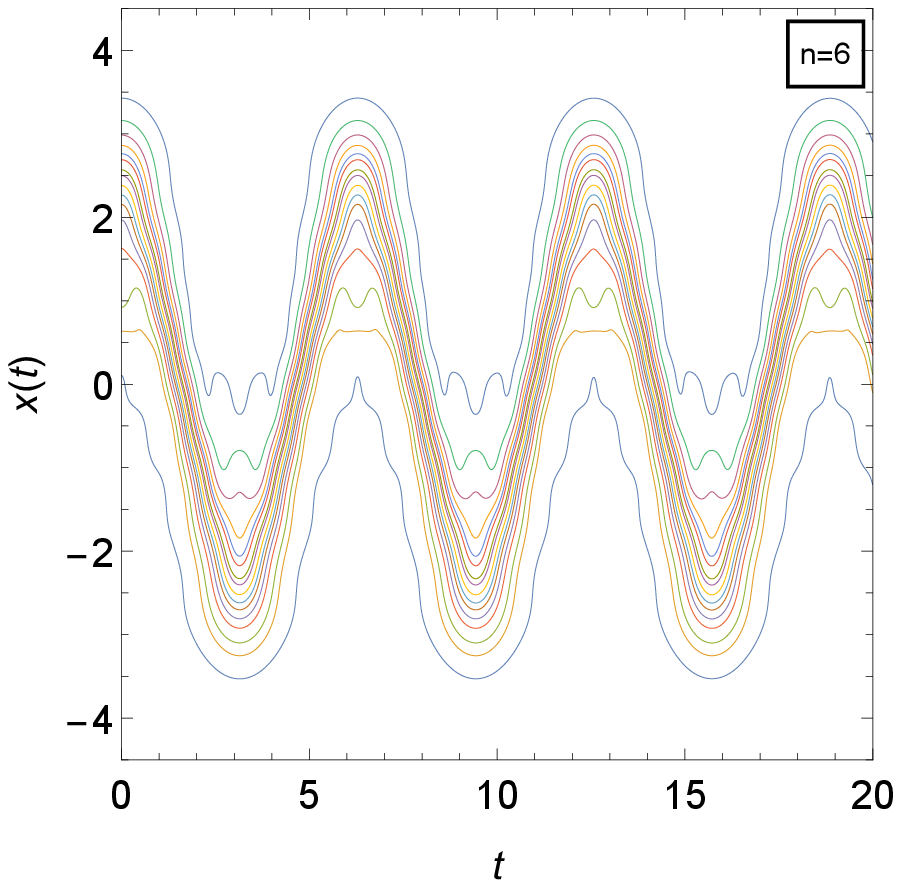} 
\vfill
 \includegraphics[width=6.cm]{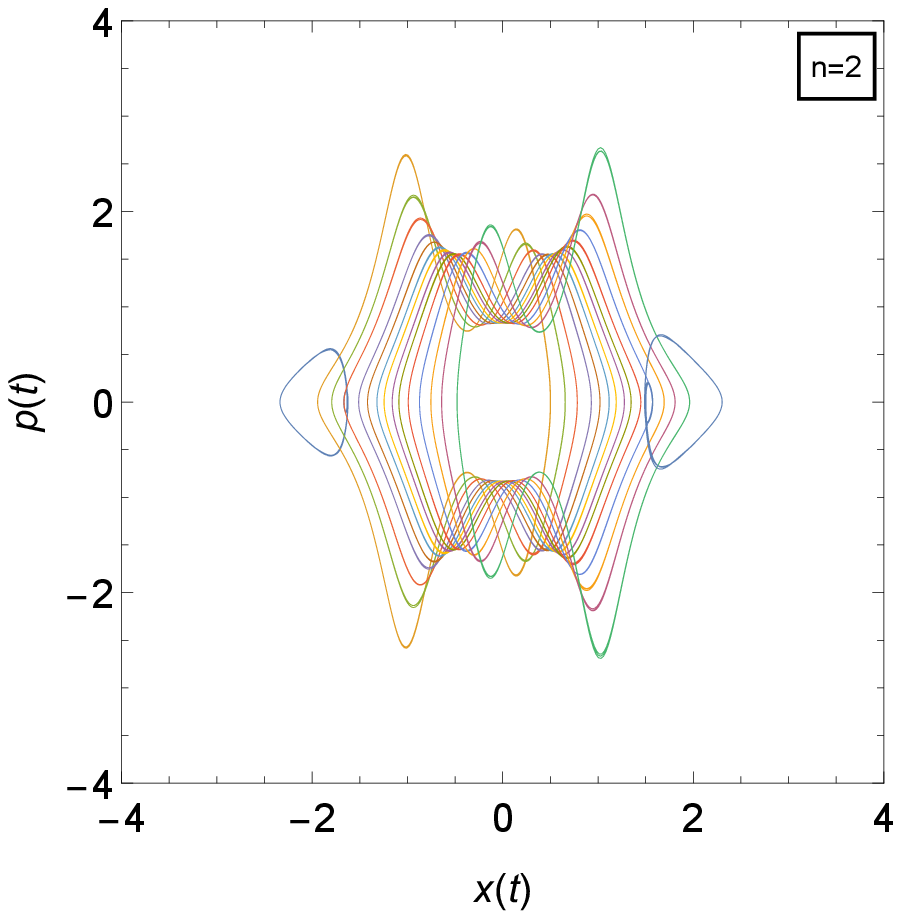}
 \includegraphics[width=6.cm]{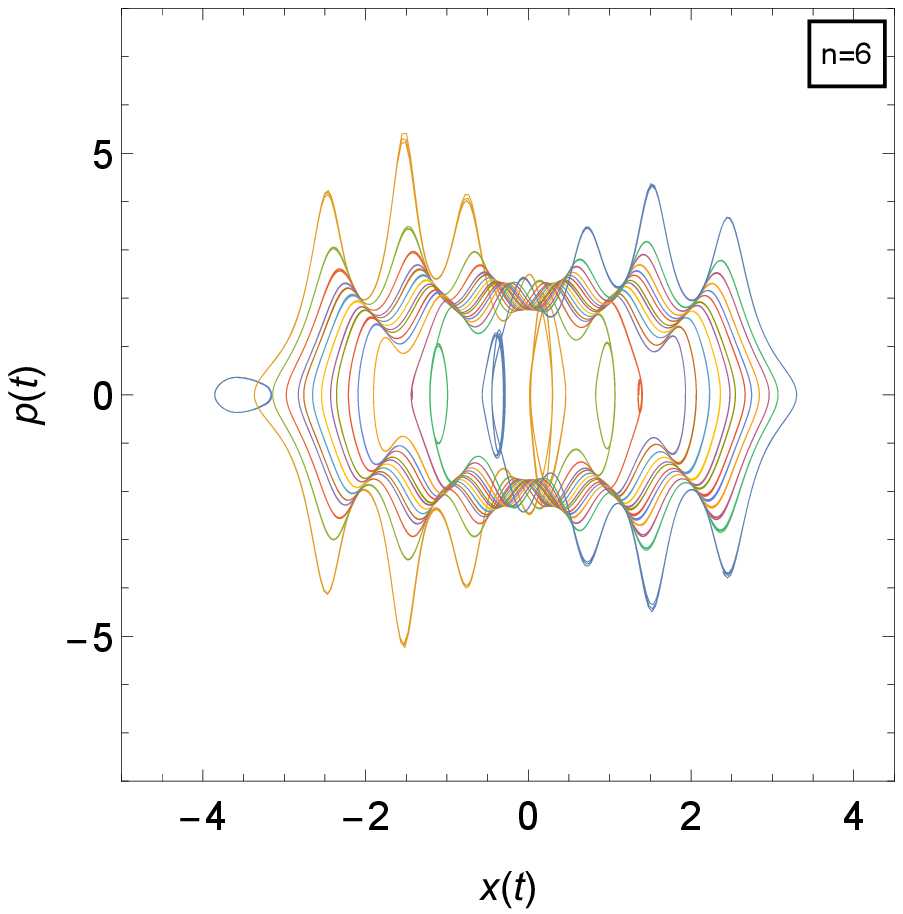}
\caption{(top) Trajectories of the quantum harmonic oscillator for the sum of eingenstates until $n=2$ and $n=6$. (bottom) The associated phase space.
} 
 \label{trajQHO}
\end{figure*}

In this case, for $n=0$ we have  static solutions, since the wave function phase $S$ does not depend on $x$. For $n>0$, on the other hand, the quantum trajectories have a non-trivial dynamics, since $\nabla S(x,t)\ne 0$. The position of the dimensionless classical harmonic oscillator is given by $x(t)=A\cos(t + \phi)$, so  the phase space obtained by the energy has concentric circles of radius $A$ as surface levels. The  trajectories are represented in FIG. \ref{trajQHO} for $n=2$ and $n=6$. Once we are dealing with a periodic system, the  phase spaces are closed and periodic.  

\begin{figure}
 \centering
 \includegraphics[width=7cm]{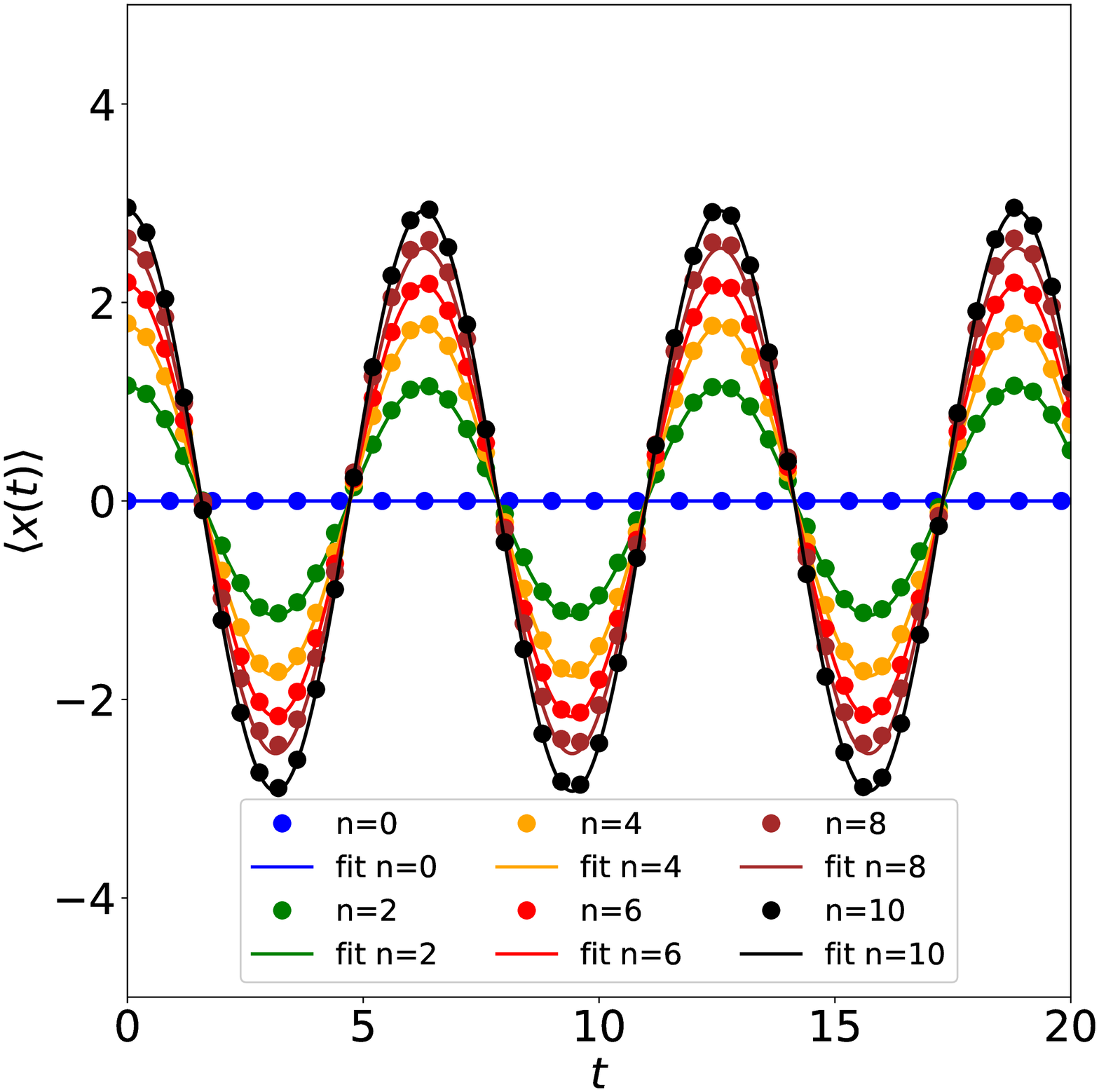}
 \includegraphics[width=7cm]{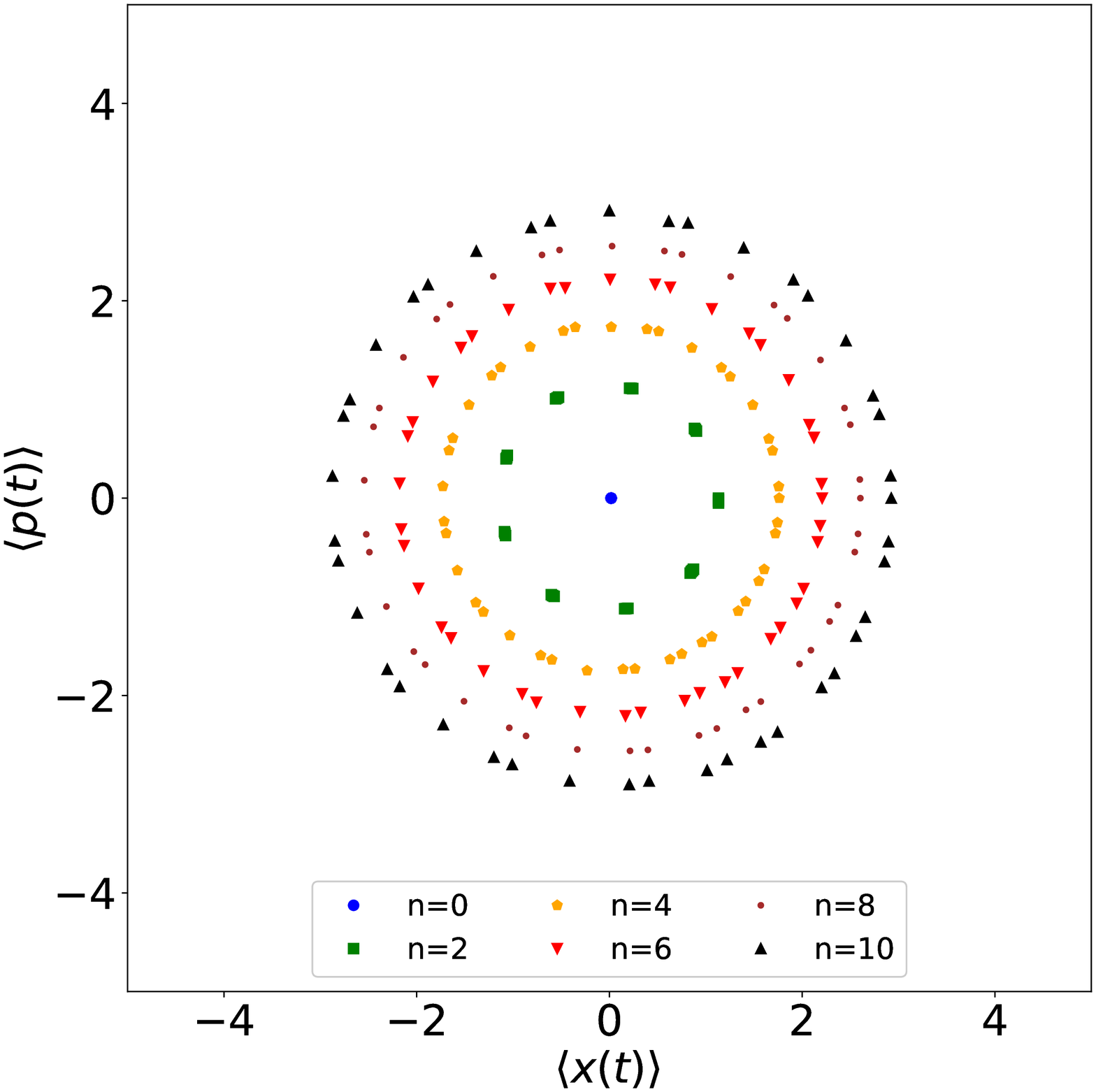}

 \caption{(left) Dynamics of $\langle x(t)\rangle_{dBB}$ for $F(t)=0$, where the solid lines are the best fits. The curves have a sinusoidal form with amplitude depending on the number of eingenstates considered in the initial superposition. (right) Classical phase space obtained by our numerical procedures. 
 } 
 \label{meanQHO}
\end{figure}

\begin{figure}
 \centering
 \includegraphics[width=7.cm]{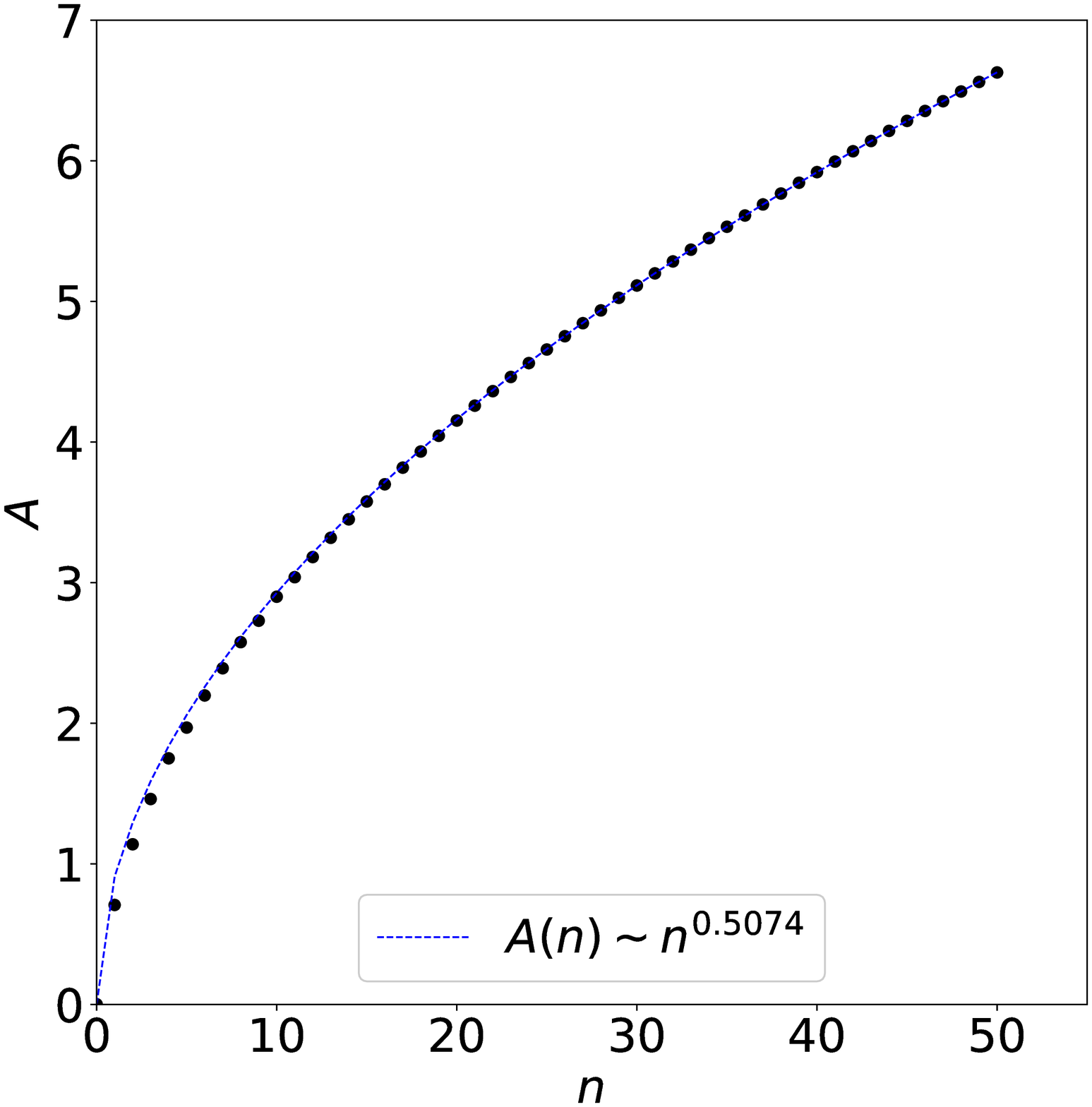}
 \caption{Plot of the classical amplitude $A_n$ versus the number of total states $n$.  The amplitude $A_n$ is in accordance  to a power law of the type $A_n\sim n^{1/2}$. We perform $A_n$  from  $n=0$ to $n=50$. 
 } 
 \label{An}
\end{figure}

Computing the average positions, we find that the best fit is given by  $\langle x \rangle_{dBB}=A_n \cos{(\omega_0 t)}$, which is very consistent  with the classical curves (see FIG. \ref{meanQHO}). The amplitude $A_n$ depends on the initial wave function and is exactly $\langle x_0\rangle_{dBB}$, in such way that as $n$ grows, the amplitude of the mean oscillations also grows. To be more precise, $A_n \sim \sqrt{n}$, according to FIG. \ref{An}. As a result, considering more initial states, more energy is available for the system. Therefore, the phase space volume  increases linearly with $n$, because the classical energy is given by $E_{class}=\frac{1}{2}m \omega_0^2 A_n^2 \propto n$. The phase space volume $\Gamma$ can be easily calculated, in dimensionless variables, by the ellipses area $\Gamma=\pi A_n^2=\pi \zeta^2 n$, where $\zeta$ is the proportionality constant.  Thence, thermal properties can be estimated from these results, for instance the classical internal energy $U(T)$. By the canonical ensemble, the partition function can be written as follows

\begin{equation}
    Z=\zeta^2\pi\int_{0}^{\infty}\,dn\,e^{-\beta \zeta^2 n/2}=\frac{2\pi}{\beta},
\end{equation}
assuming a continuum sum of states.
 The internal energy $U(T)=-\frac{\partial \ln{Z}}{\partial \beta}$ yields $U(T)=k_BT$, in agreement with the equipartition theorem. Even whether we consider a discrete number of states for $E_{class}$ in the canonical ensemble , we have the same result. Indeed,
\begin{equation}
    Z=\sum_{n=0}^{\infty} e^{-\beta \zeta^2 n/2}=\frac{1}{1-e^{-\beta \zeta^2/2}}\approx \frac{2}{\beta \zeta^2}\,,
\end{equation}
where, by taking $\beta\to 0$ and using the definition of $U(T)$, we have  $U(T)\approx k_BT$, as expected.

\subsection{Constant force}

\begin{figure}
 \centering
 \includegraphics[width=6.9cm]{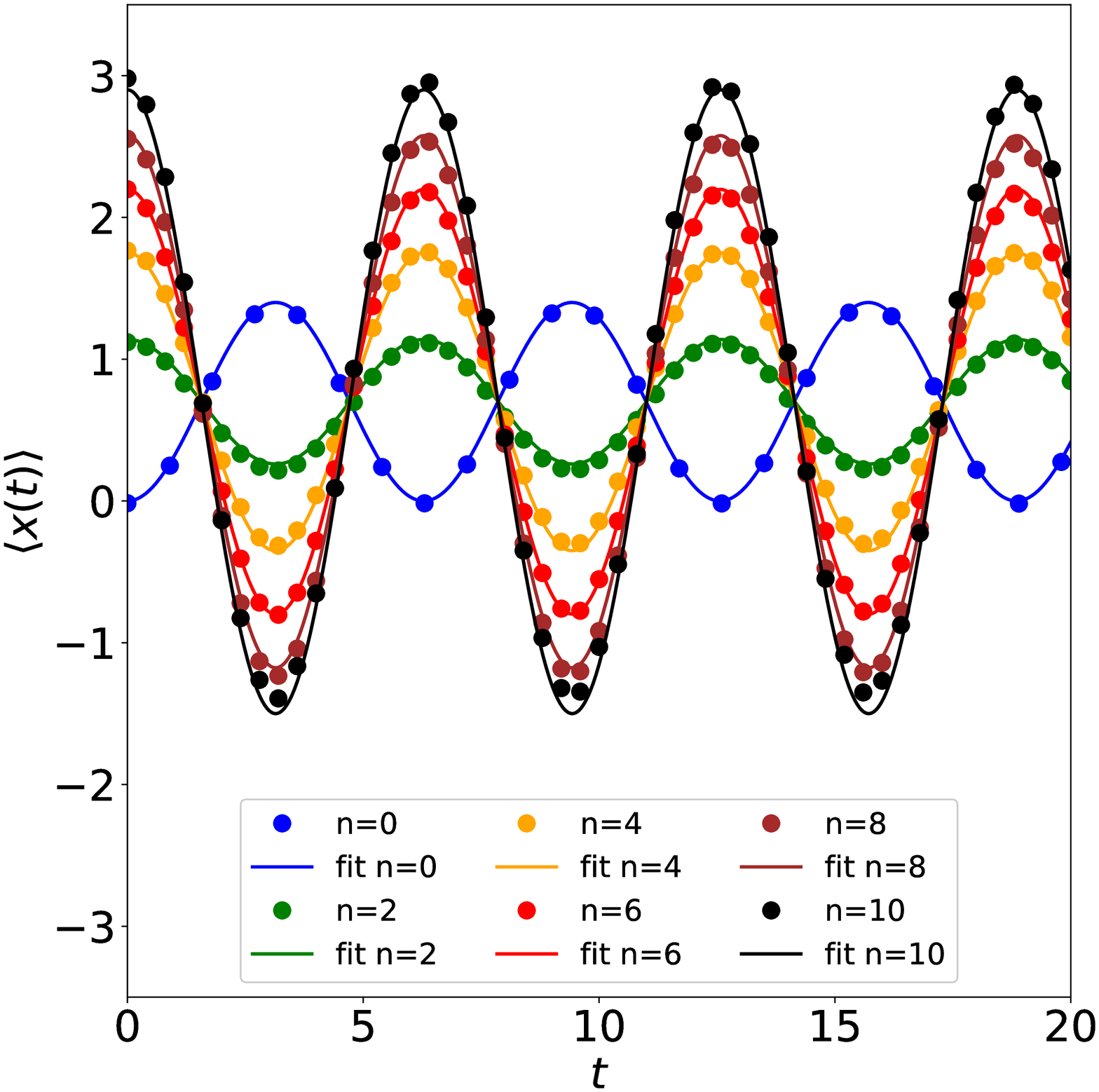}
 \includegraphics[width=7cm]{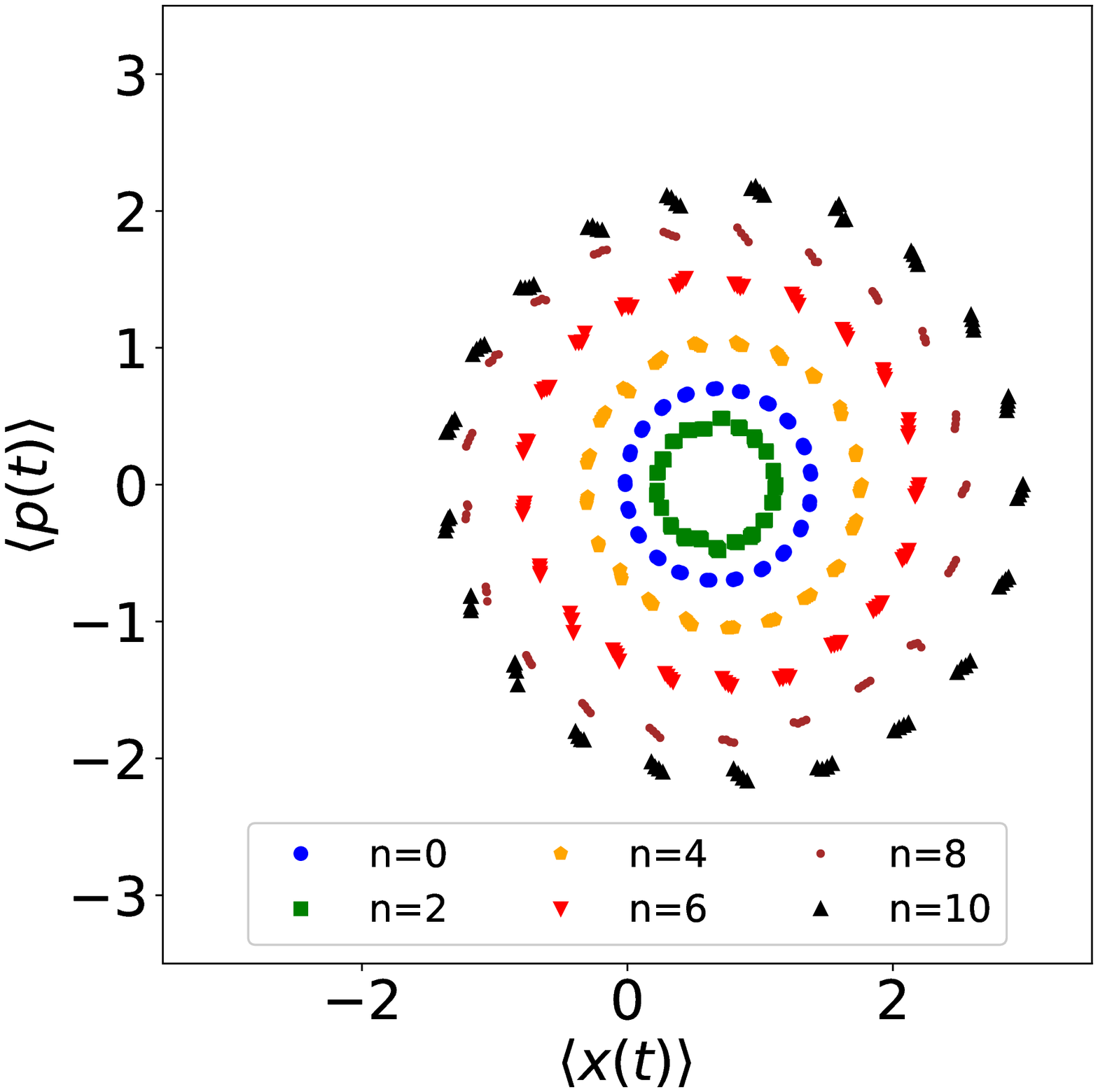}
 \caption{(left) Average trajectory $\langle x(t)\rangle_{dBB}$ for a constant force $F(t)=0.7$. (right) Respective classical phase space obtained numerically. Note the change in amplitude and the shift in the position around the oscillations.} 
 \label{constQHO}
\end{figure}

Let us focus on the constant force case. Despite of its simplicity, this example  can be used, for instance, to simulate a charged harmonic oscillator in a uniform  external electric field, and the effect of a weak gravitational field as well. We set, without loss of generality, $F(t)=0.7$. We notice that the general solution for the mean positions is given by  $\langle x(t) \rangle_{dBB}= \langle x(t)\rangle_{dBB}^{HO}+0.7(1-\cos(t))$, as expected (see FIG. \ref{constQHO} ). The classical solution is exactly the same of the previous one, despite of the translation of  the position and amplitude by the value of  0.7. For an arbitrary constant, we must have $x\to x+C$ and $A\to |A-C|$. The mean trajectories reproduce such property. Furthermore, in contrast with null force case, now for $n=0$ the particles have non-static trajectories and oscillatory averages, since the initial positions are not distributed around the equilibrium point. However, the momentum remains the same, implying that the phase space is shifted in the $x$ direction.

\begin{figure*}
 \centering
 \includegraphics[width=6.cm]{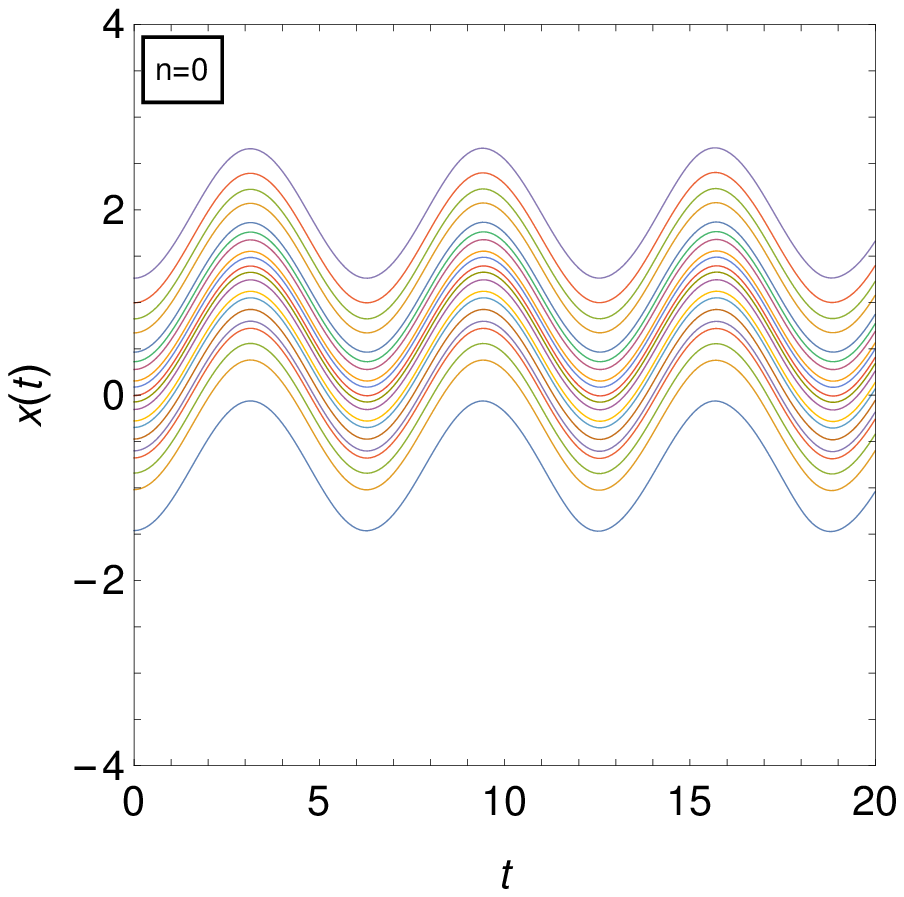}
 \includegraphics[width=6.cm]{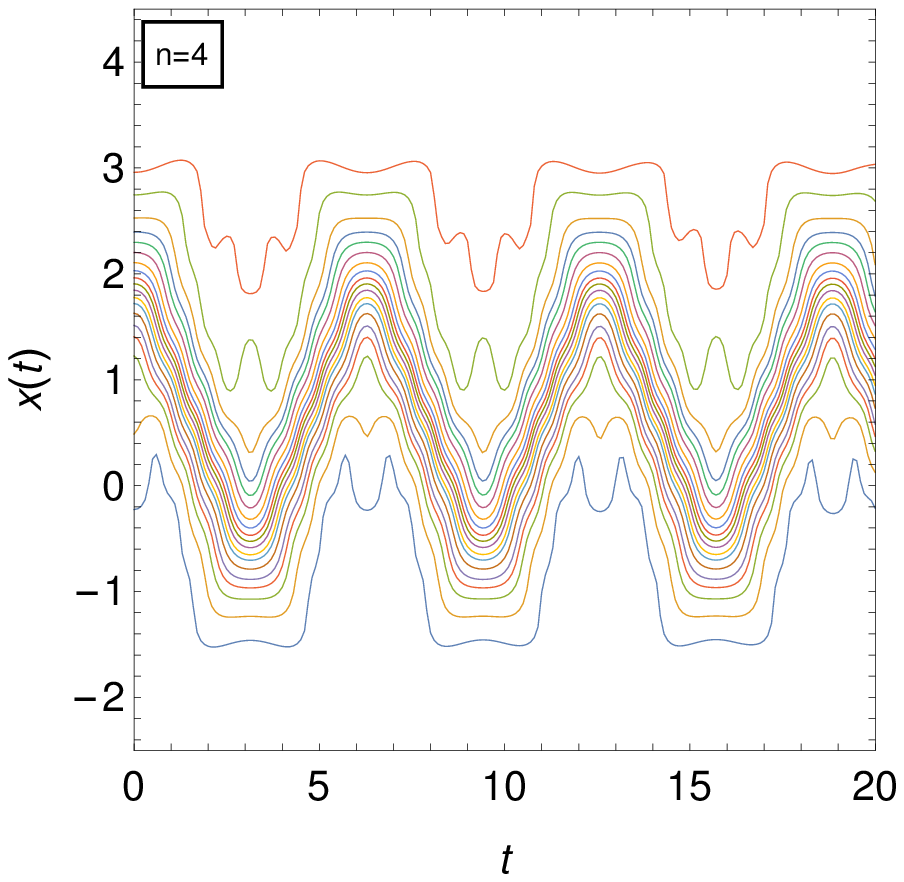} 
\vfill
 \includegraphics[width=6.2cm]{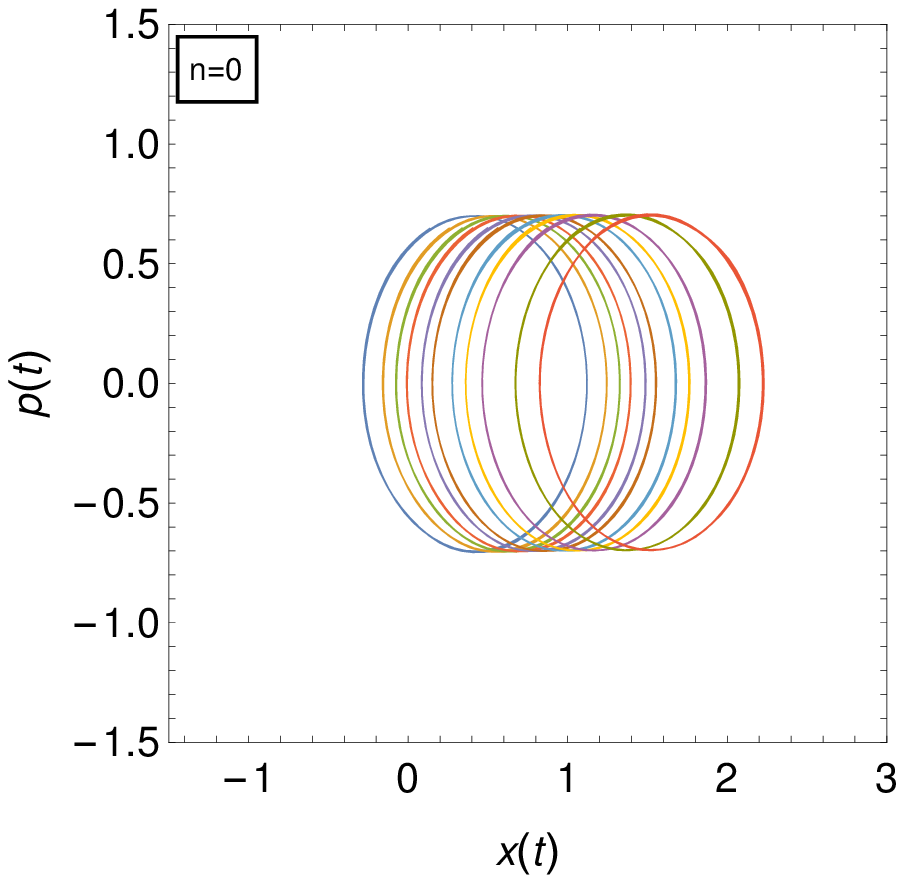}
 \includegraphics[width=6.cm]{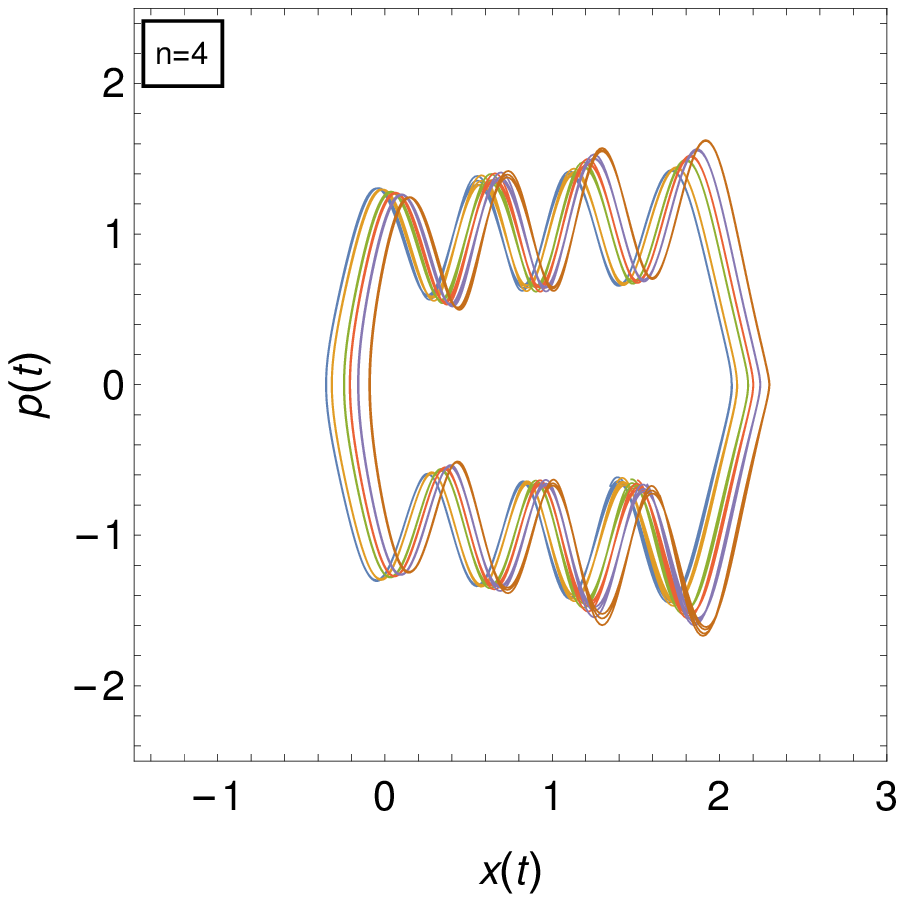}
\caption{(top) Trajectories of the quantum harmonic oscillator subject to an external constant force $F=0.7$, for the sum of eingenstates until $n=0$ and $n=4$. (bottom) The associated phase space for central trajectories (close to $\max{\{|\Psi(x,t)|^2\}}$). Note that the trajectories are shifted 
in the $x$ direction.} 
 \label{trajQHOFconst}
\end{figure*}

\subsection{Impulsive Force}

\begin{figure*}
 \centering
 \includegraphics[width=6cm]{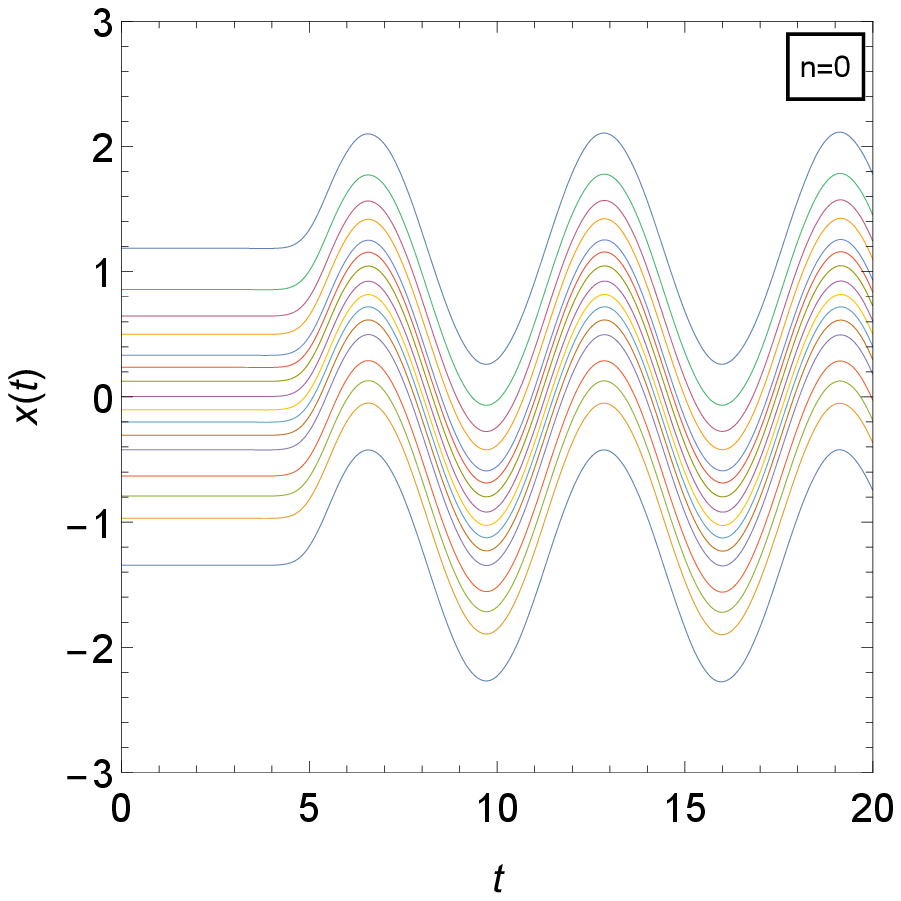}
 \includegraphics[width=6cm]{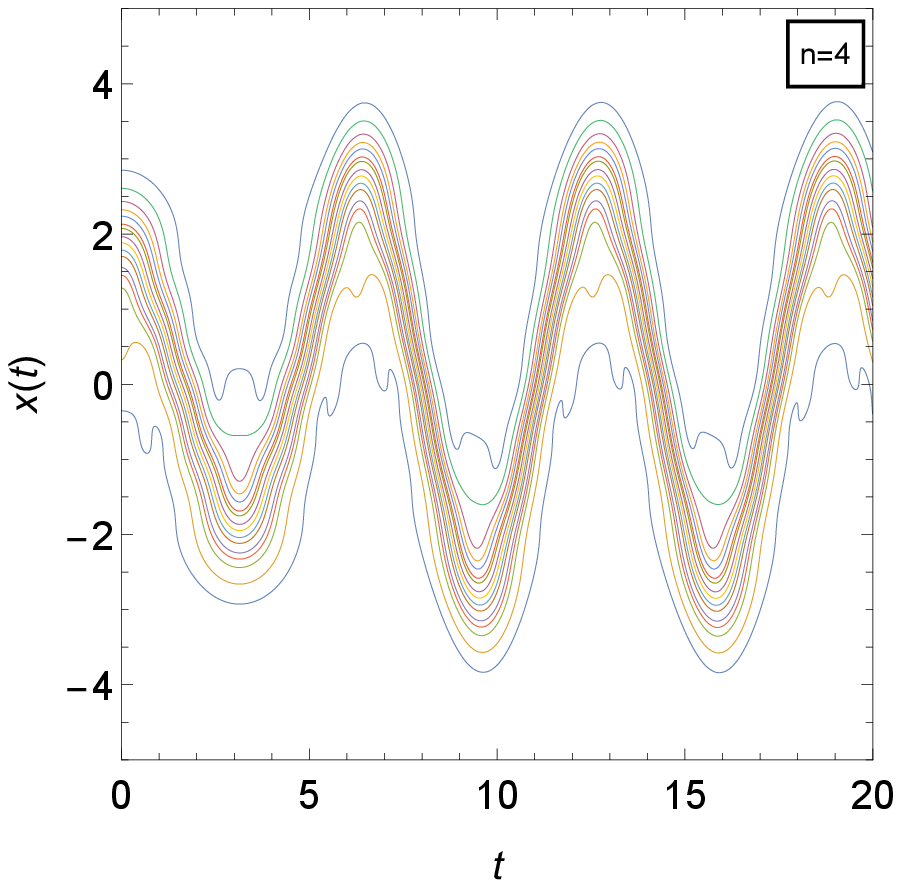} 
\vfill
 \includegraphics[width=6.2cm]{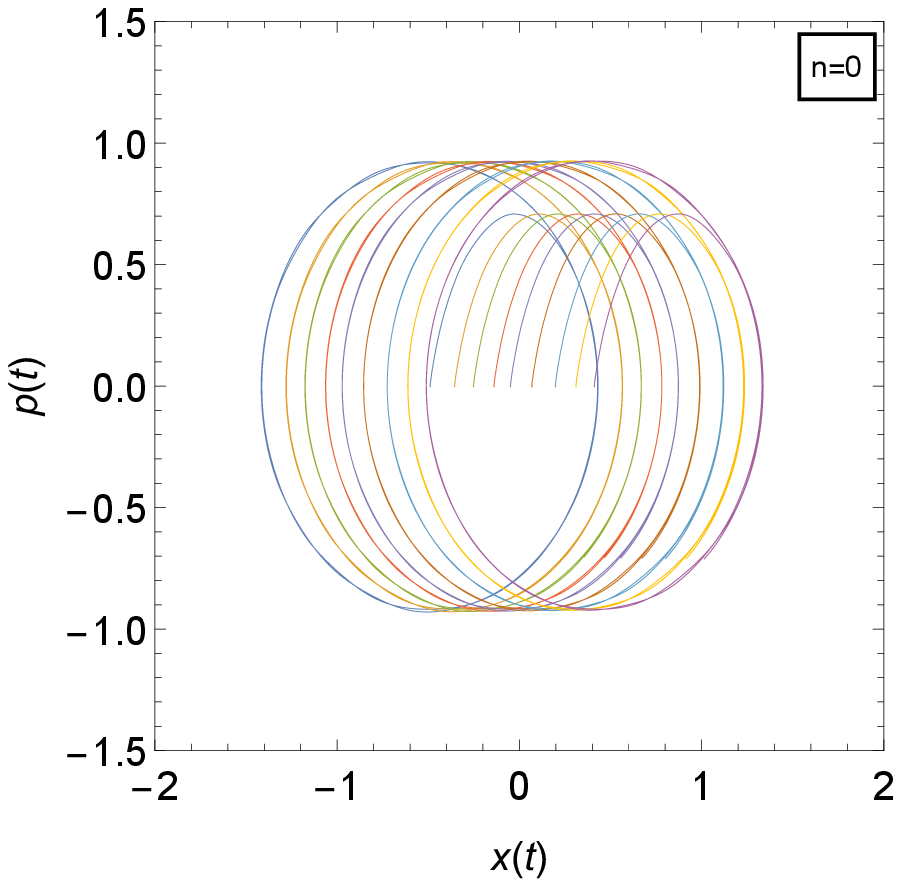}
 \includegraphics[width=6cm]{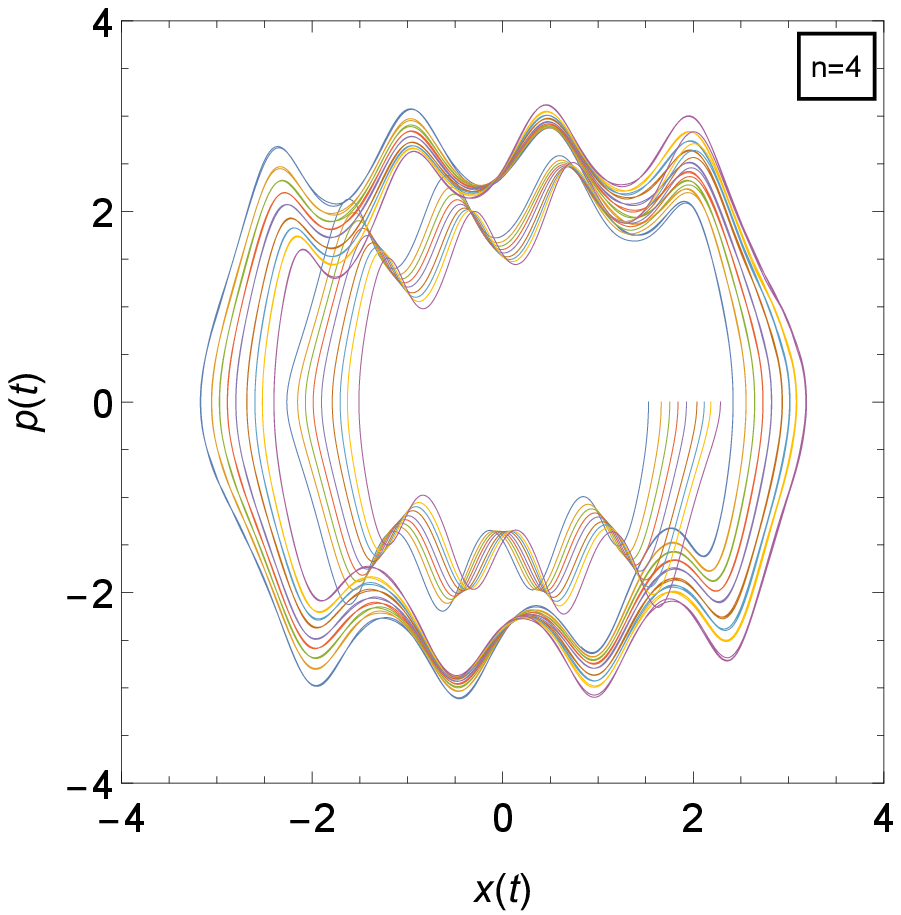}
\caption{(top)Trajectories of the impulsive case. (bottom) Phase space for central trajectories (close to $\max{\{|\Psi(x,t)|^2\}}$). Two asymptotically regimes can be observed, before and after the impulse.} 
 \label{trajgauss}
\end{figure*}

The impulsive forces have a wide class of applications in quantum systems. As examples we can cite non-adiabatic transitions  \cite{CokerXiao1995}, optomechanics \cite{VitaliMancini2001, BennetBowen2018} and prediction  of Gaussian quantum systems \cite{HuangSarovar2018}. In our simulations we consider a fast acting impulsive force, modeled by $F(t)=\frac{1}{\sqrt{2\pi\sigma^2}} \exp{\left(-\frac{(t-t_{\mu})^2}{2\sigma^2}\right)}$, were we set the values of $t_{\mu}=5$ and $\sigma=0.4$. The asymptotic initial and final states are just the quantum harmonic oscillator, studied in the previous case. The difference is that the force brings more energy to the system, exciting more eingenstates and changing each individual weight of the linear superposition. The effect of such perturbation can be viewed in FIG. \ref{trajgauss}. Before the force acts, we have the same trajectories than the unforced case, passing to present a different behavior close to the peak of the Gaussian at $t=5$. For $n=0$, for example, the trajectories that are initially static pass to oscillate after the force ceases. As the effect of this perturbation, the volume of the phase space increases.

\begin{figure}[htb]
 \centering
  \includegraphics[width=7cm]{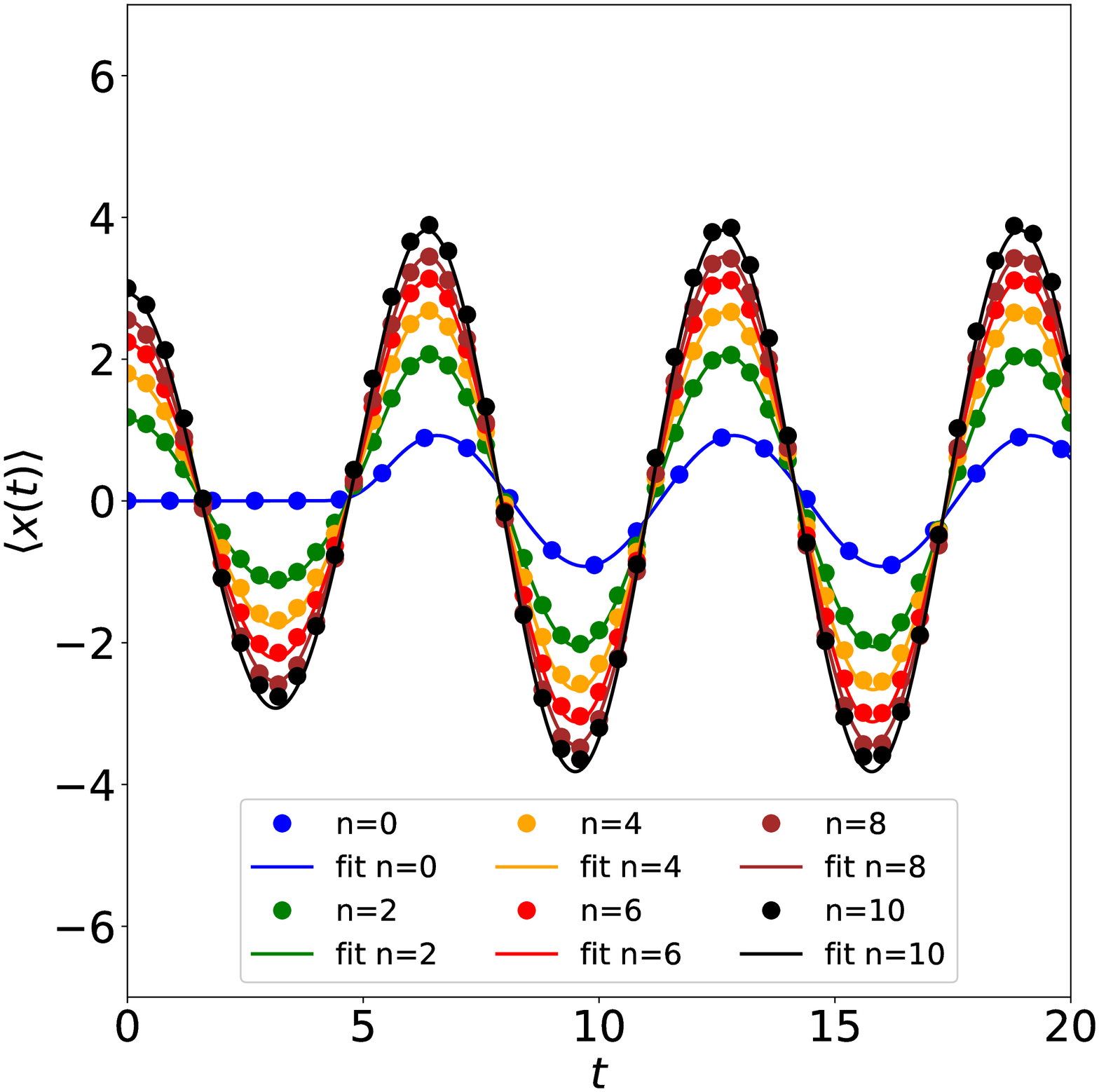}
 \includegraphics[width=7cm]{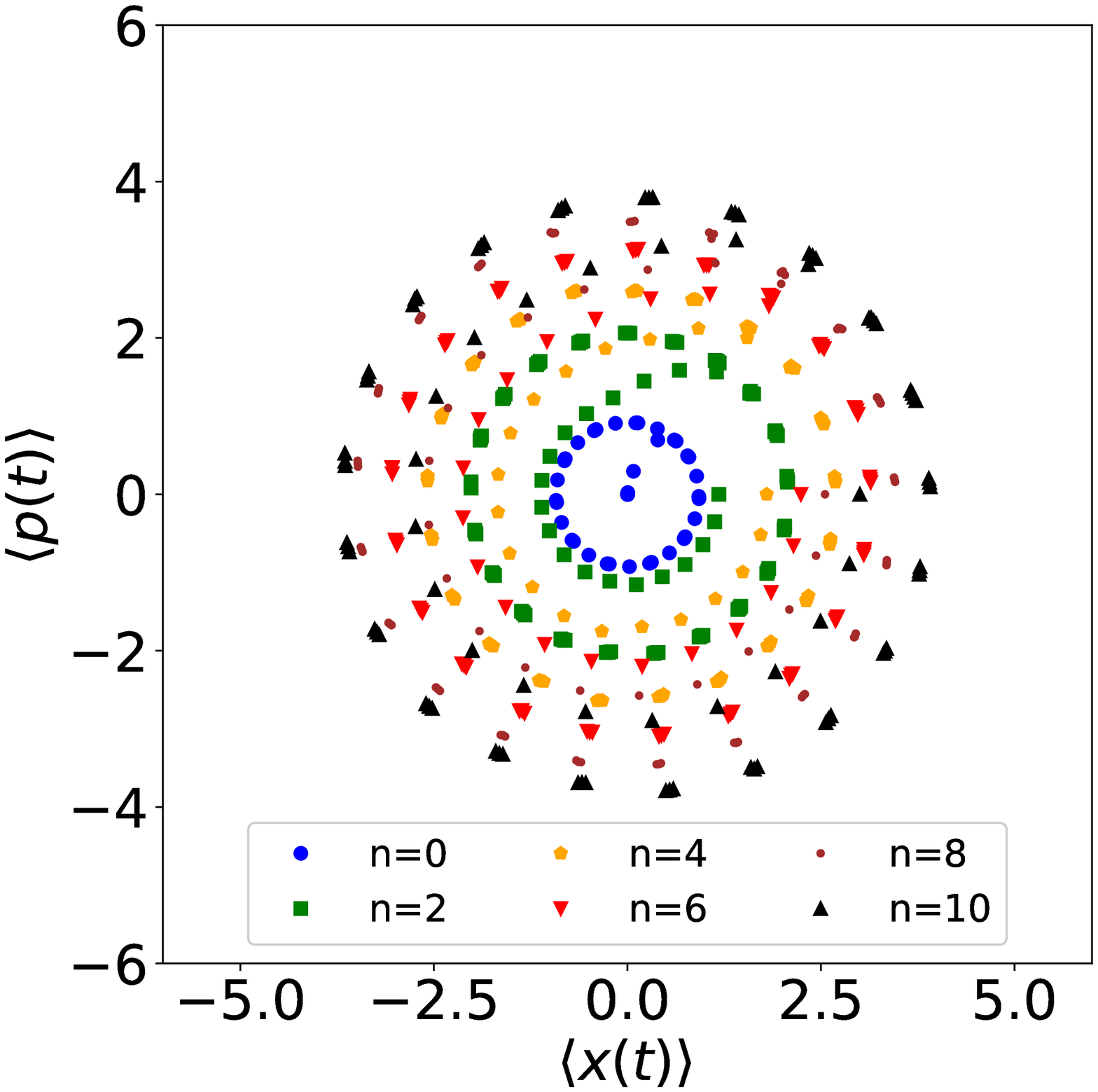}
 \caption{(left) Dynamics of $\langle x(t)\rangle_{dBB}$ for the impulsive force. (right) Associated classical phase space obtained by our numerical procedures. After the impulse at $t=5$ the amplitude of the oscillations and the radius orbit increases, as a result of the interaction.
 } 
 \label{meangauss}
\end{figure}

The averages are illustrated in FIG.~ \ref{meangauss}. For all considered values of $n$ , the initial and final states have a sinusoidal form with different amplitudes, increasing  the mean energy of the ensemble. This can be best understood looking to the phase space. In all examples we start in a circle of specific radius and, when the force starts to become relevant we are rapidly induced into a larger radius orbit, remaining there after the force stops. As we also expect, the averages obey  Eq.~ \eqref{conv}.

\subsection{Sinusoidal force}

\begin{figure}
 \centering
 \includegraphics[width=7cm]{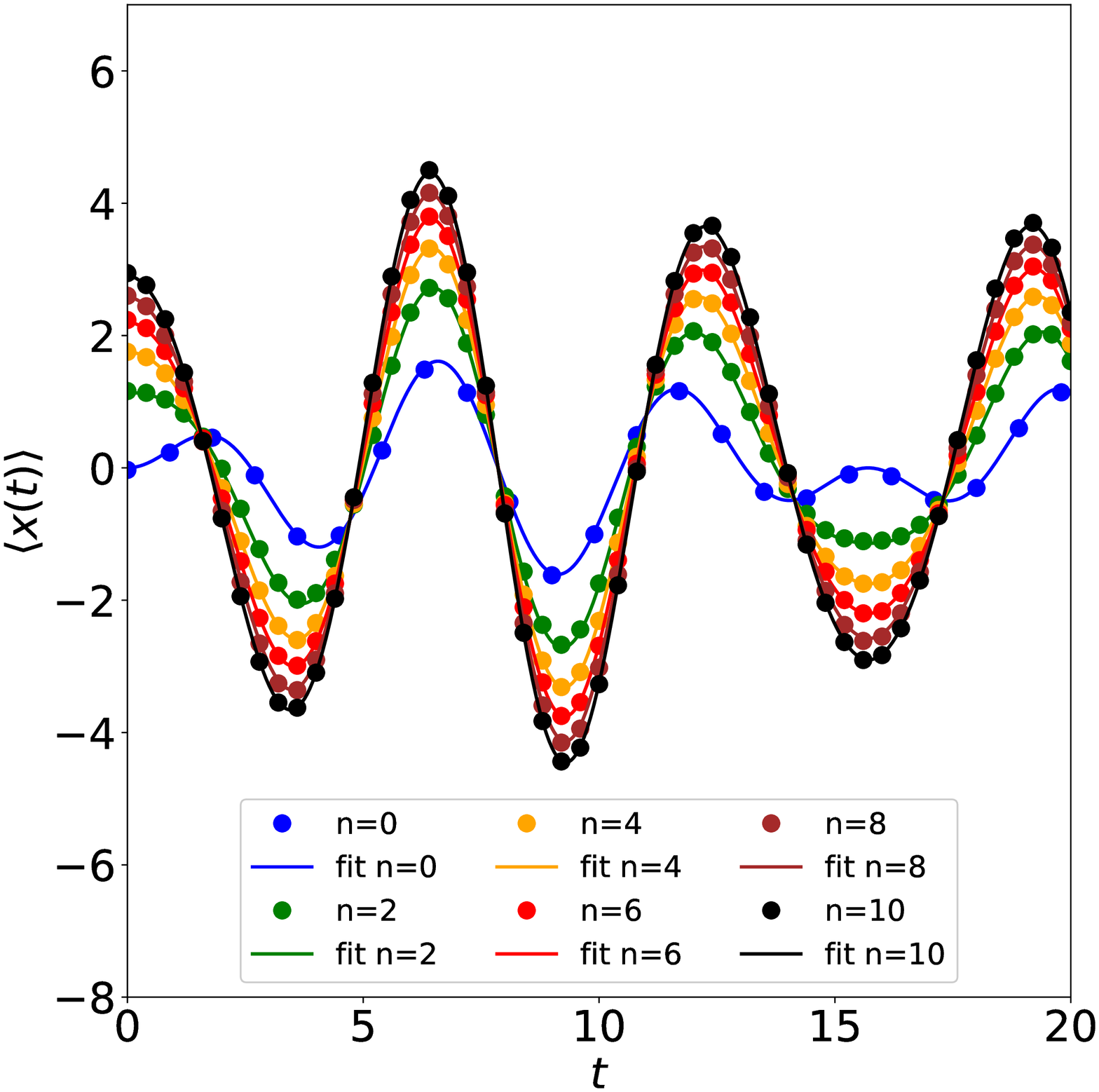}
 \includegraphics[width=7cm]{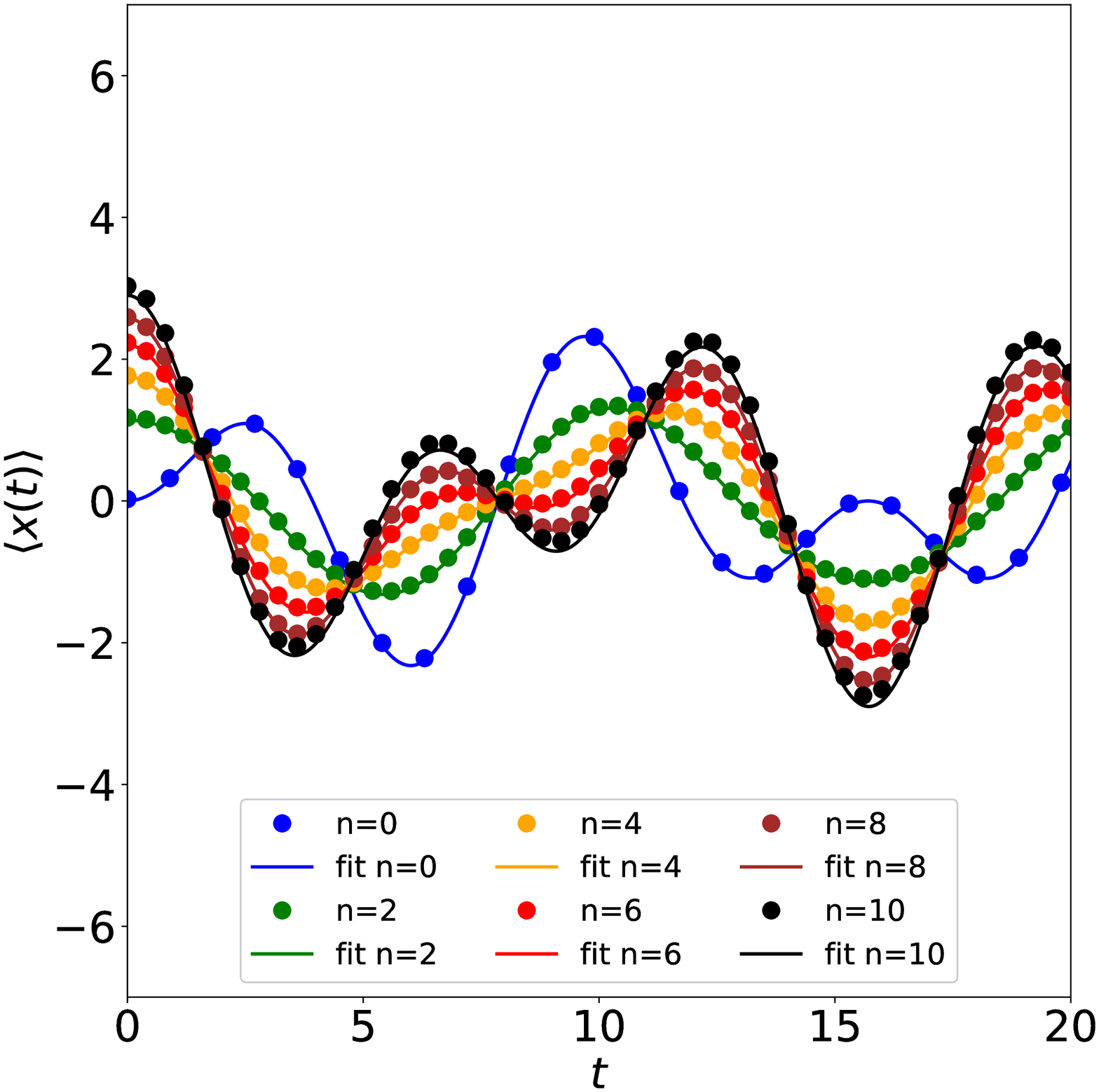}
 \caption{Dynamics of $\langle x(t)\rangle_{dBB}$ for the sinusoidal force. Two frequencies are considered: $\Omega=1.4$ (left) and $\Omega=0.6$ (right). 
 } 
 \label{meanhighlow}
\end{figure}

Let us focus on the sinusoidal  force $F(t)=F_0\cos{(\Omega t)}$, where $t$ is the dimensionless temporal variable and $\Omega$ is the ratio between the frequency $\omega$ of the external force and the fundamental frequency $\omega_0$, this is to say, $\Omega=\omega/\omega_0$. The values of $\Omega$  for the high and low frequency cases are chosen as $\Omega=1.4$ and $\Omega=0.6$, respectively, with an amplitude of $F_0=0.8$. We anticipate that their averages obey, with high precision, the classical laws
(see FIG.~\ref{meanhighlow}).

\begin{figure}[htb]
 \centering

 \includegraphics[width=7 cm]{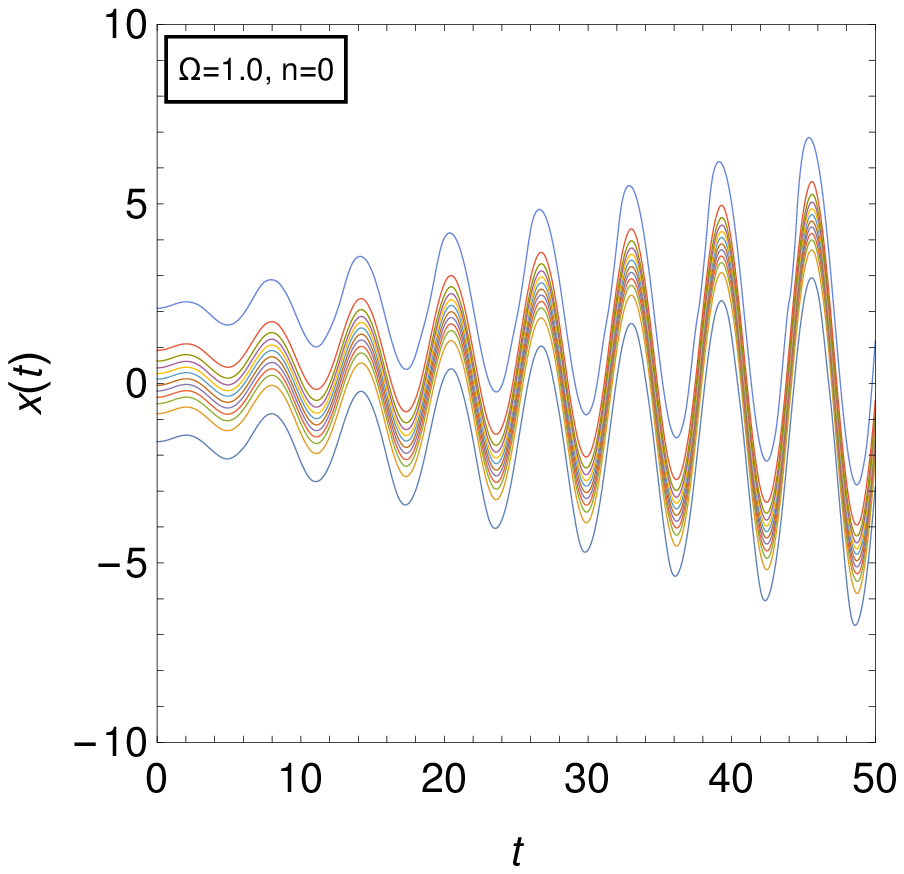} 
 \includegraphics[width=6.75cm]{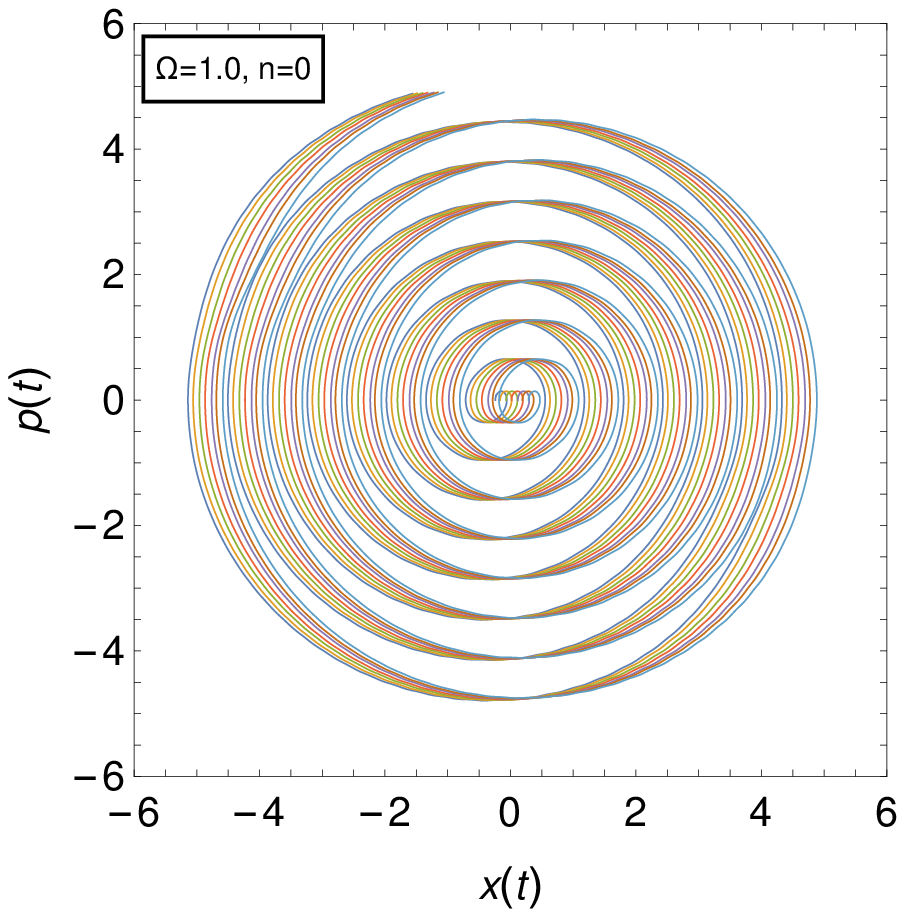}
\caption{(left) Resonant trajectories for $n=0$ with $F_0=0.2$. As time passes the amplitude of the oscillations grows, corresponding to an expanding phase space (right).
} 
\label{restraj}
\end{figure}

\begin{figure}[htb]
 \centering
 \includegraphics[width=6.75cm]{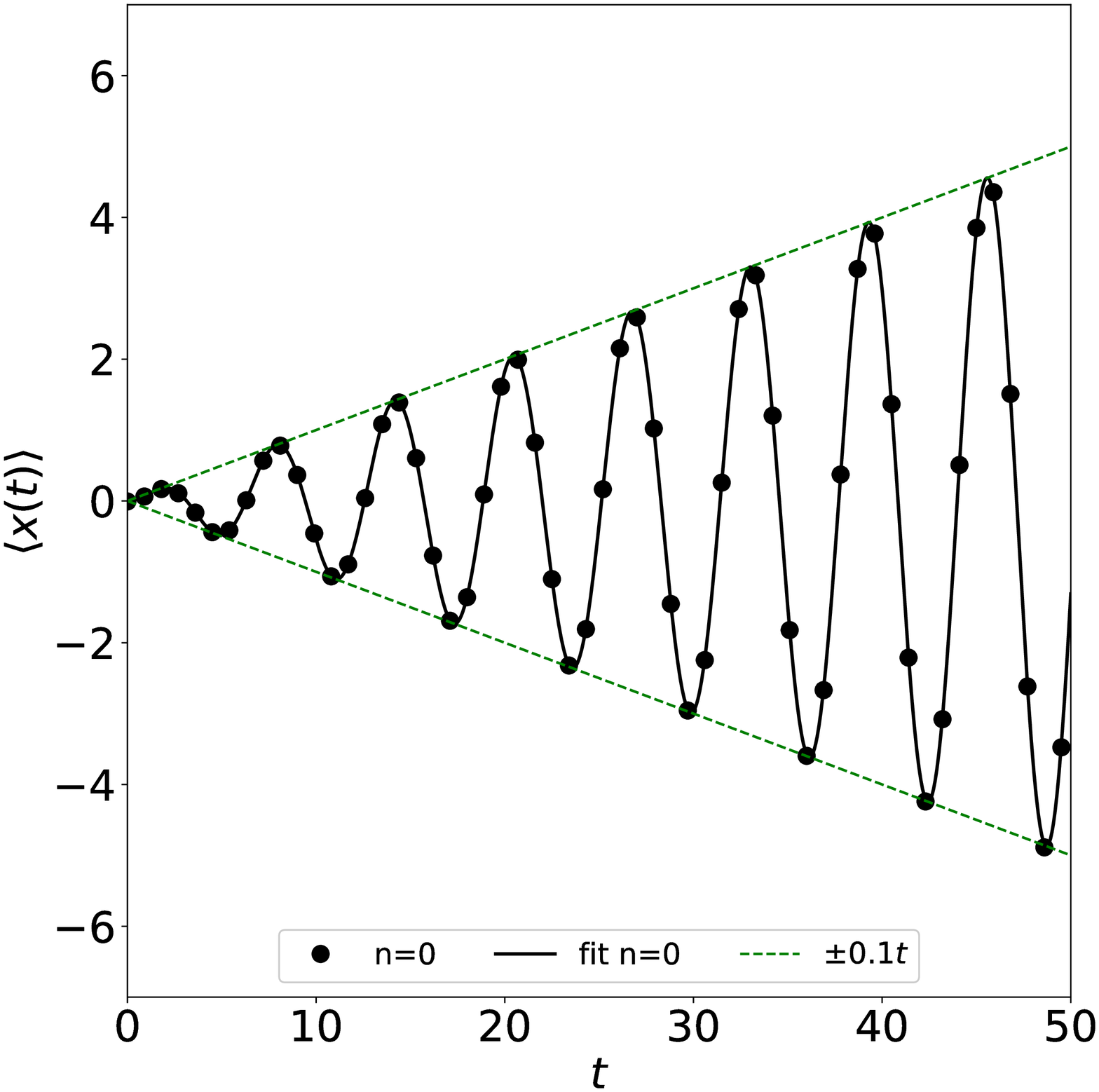} 
 \includegraphics[width=7cm]{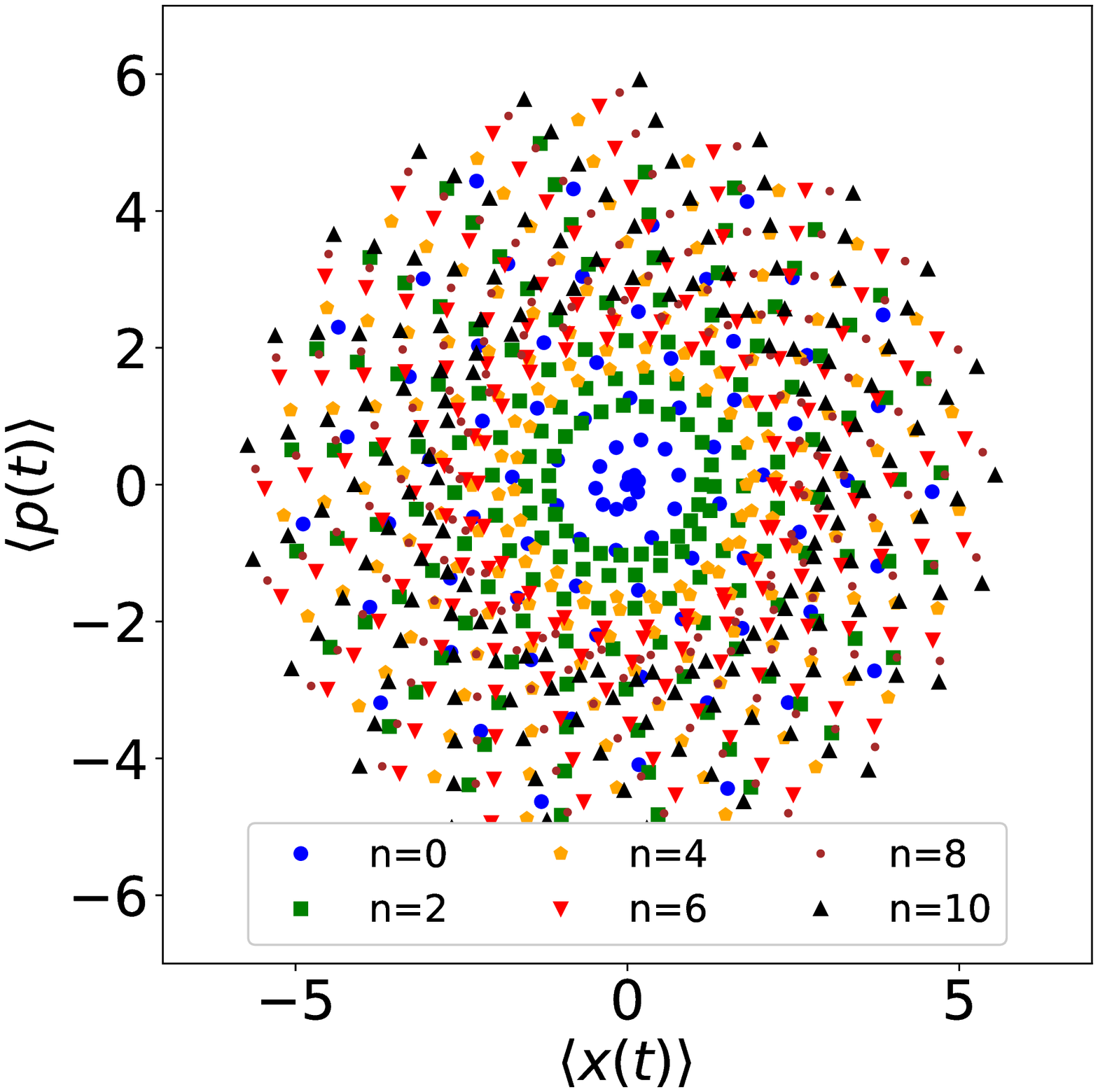}
 \caption{ (left) Time evolution of $\langle x(t)\rangle_{dBB}$ for the resonant force with $n=0$. (right) The classical phase space. The amplitude grows linearly with $F_0 t/2$, according to a classical law.
 } 
 \label{meanres}
\end{figure}

 As in the classic case, when a vibrating system is excited by a force with the same frequency, which is equivalent to take $\Omega=1$, the quantum resonance is observed . This phenomenon has been  intensively investigated in the literature,  with many applications as we can see in atom optics \cite{Fishman2002}, experimental quasi-momentum measurements \cite{DanaTaluk2008}, multichromophoric energy transfer \cite{Duqueetal2015}, electric-dipole moment experiments \cite{Silenko2017}, and   nano-resonance \cite{Egorov2022}.
We restrict ourselves to show  and to asses  the quantum phase spaces and trajectories only for $n=0$. In order to avoid the interference from unwanted boundary effects in the results, we take $L=15$ and $F_0=0.2$. The trajectories are presented in FIG. \ref{restraj}. From the phase space we can see a monotonic  increase in energy, since the trajectories spiral outwards as time passes, always expanding the volume of the phase space. We observe a typical resonant phenomenon for the mean trajectories, where the amplitude grows linearly with $F_0 t/2$ (see FIG.~ \ref{meanres}) as the force acts.
 
\subsection{Quantum Duffing oscillator and resonance}
Let us consider the quantum Duffing oscillator modeled by the following Hamiltonian
\begin{align}
    \hat{\mathcal{H}}=\frac{\hat{p}^2}{2}+\frac{\hat{x}^2}{2}+\frac{\lambda}{4}\hat{x}^4-F_0\hat{x}\cos(\Omega t),
\end{align}
which can be viewed as a driven anharmonic oscillator  with $\Omega$  defined  as the ratio between the external force frequency $\omega$ and the natural frequency $\omega_0$ of the correspondent harmonic oscillator obtained by taking $\lambda=0$. This model corresponds to a particular case of the Duffing equation \cite{Duffing1918}, which finds applications in classical and quantum systems such as stiffening springs, beam buckling, nonlinear electronic circuits, superconducting Josephson parametric amplifiers, and ionization waves in plasmas \cite{Hans2008}. In order to compare the results with the resonant case in the previous subsection, we consider $n=0$, $F_0=0.2$, $\Omega=1$ and for the quartic potential responsible for the anharmonicity $\hat{V}_{anh}=\frac{\lambda}{4}\hat{x}^4$  we set $\lambda=0.01$ as a small perturbation. 

\begin{figure*}[htb]
 \centering
 \includegraphics[width=7.5cm]{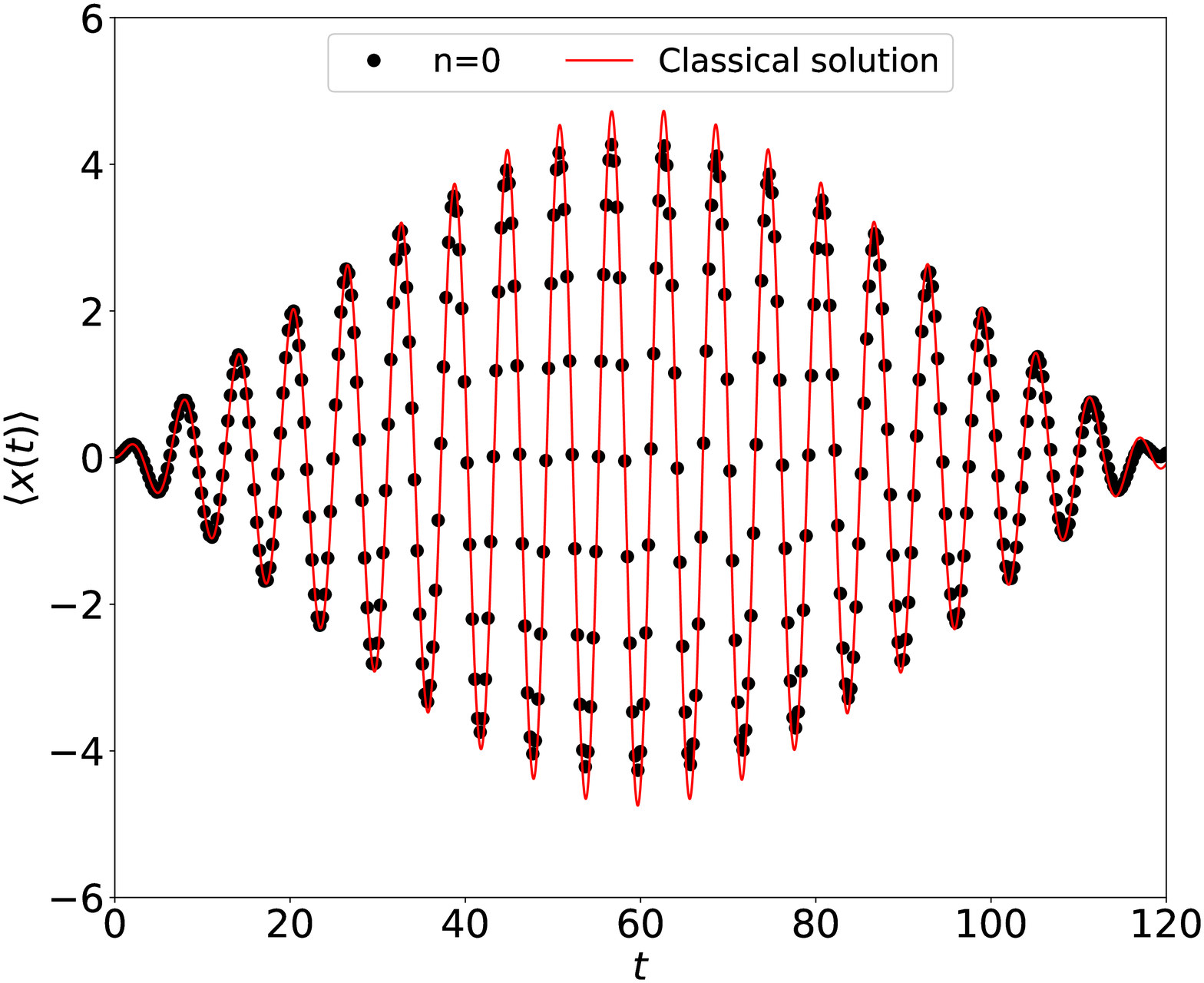} 
 \includegraphics[width=7.5cm]{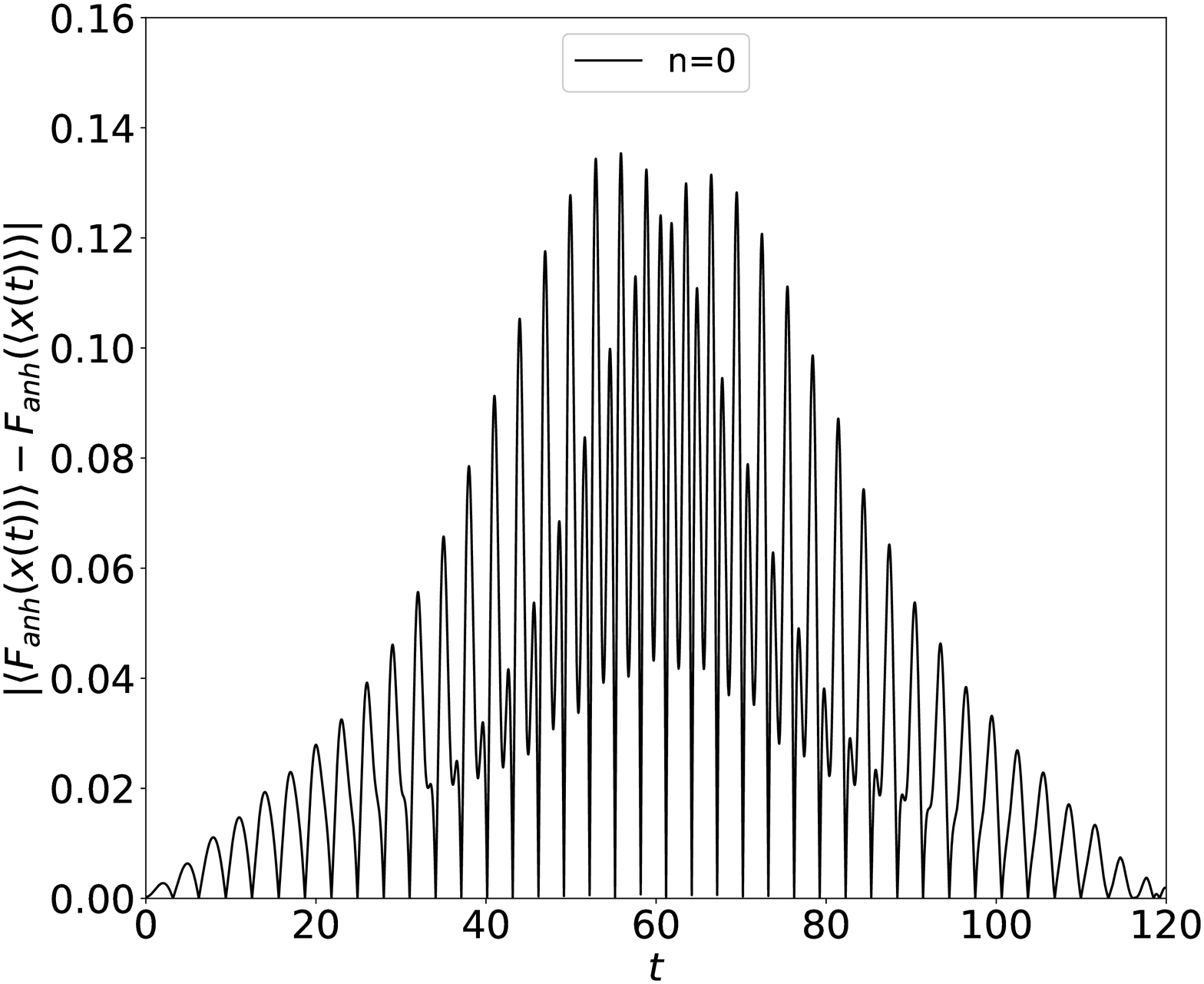}
\caption{(left) Time evolution of $\langle x(t)\rangle_{dBB}$ for the quantum Duffing oscillator (black points) and for the numerical solution of its classical analogue (red line) with initial conditions $(x_0,p_0)=(0,0)$. (right) Difference between $\langle F_{anh}(x(t))\rangle_{dBB}$ and $F_{anh}(\langle x(t)\rangle_{dBB})$ where $F_{anh}(x)$ is the anharmonic force, namely $-\lambda x^3$.
} 
 \label{Duf}

\end{figure*}
In this case, a non-trivial difference between  the first and third momentum, $\langle x(t)\rangle_{dBB} $ and $\langle x(t)^3\rangle_{dBB}$, emerges as a consequence of the nonlinear force, namely $F_{anh}(x)=-\lambda x^3$, as in FIG. \ref{Duf}. The difference $\langle x(t)^3\rangle_{dBB} -\langle x(t)\rangle^3_{dBB}\neq0$ makes the quantum result slightly different from the classical one. Even for a small perturbation constant, as the values of the coordinate become large, this difference also increases. This result is naturally expected since the Ehrenfest theorem corresponds to the classical analogue only for linear and constant forces. The Ehrenfest theorem in the form of Eqs.~\eqref{Ehrenfest dBB1} and \eqref{E} is still valid, however it does not correspond to an exact classical solution, but an approximate one. Let us emphasize that the average of the trajectories shown in FIG. \ref{Duf} are in good agreement with the classical solution, evidencing that, at small perturbations, the correspondence between the classical and quantum equations still holds. At the beginning, when $x$ is small, the quadratic potential prevails. As a result, the quantum trajectories are subject to a resonant force that increases the amplitude of the oscillations as time passes. This leads the trajectories far away from the origin, where the effects of the anharmonic potential start to become relevant. Instead of having a constantly increasing amplitude as in the previous subsection, we have a beat-like pattern, since there is only a slight difference between the external frequency $\omega=\omega_0$ and the fundamental frequency $\omega_0+\epsilon$, where $\epsilon$ is a small correction due to the anharmonic perturbation.

We would like to point out that for more complicated potentials as Coulomb or Morse potential, among others, which can not be accurately  described by polynomials, the Ehrenfest theorem remains valid and can be studied by this numerical approach. However, focusing in classical-quantum equivalence of the trajectories, we concentrated on small corrections of the quantum harmonic oscillator and linear response theory, since this equivalence is achieved exactly ($\lambda=0$) or almost exactly ($|\lambda|\ll 1$) for these cases. 

\section*{ Conclusions}

In this work we prove numerically that the Ehrenfest theorem is also valid in the Bohmian mechanics, even for a certain non-conservative system as the forced quantum harmonic oscillator and Duffing oscillator. Since our approach is quite simple and has a few number of derivatives, it allows us to investigate a wide class of quantum systems without compute and simulate non-linear partial differential equations such as Eqs. \eqref{Hamilton-Jacobi} and \eqref{continuity}. This method possibly can be a simpler manner to achieve classical results with very small boundary effects, at least for simple geometries. We observe that, as we increase the number of states $n$, the quantum trajectories become more and more different of the classical analogues, however the averages obeys a classical law given in terms of the convolution \eqref{conv}, and the amplitudes of the oscillations depend on the number of states in the initial conditions. For $F=0$ we obtain that $A_n \sim \sqrt{n}$ and the classical energy linearly depends on $n$ as well as its  classical phase space volume. Thence, we reproduce the equipartition theorem by taking $n$ as a coordinate. Additionally, we can raise a question about the correspondence principle, since in the resonant case in subsection 5.4, the quantum trajectories preserved the structure of the classical result without taking account their average, and curiously, their time evolution was preserved even for non-zero values of $\hbar$.

For the Duffing oscillator, the Ehrenfest theorem can provide a satisfactory result for small perturbations when we performed the trajectories average, becoming  a good starting  point to explore other types of nonlinear potentials, even in the cases where the quantum-classical equivalence is not satisfied. Therefore, we conclude that the numerical procedures presented here can be used to investigate other quantum models where high accuracy and precision are necessary, such as integrable systems exhibiting unstable orbits, chaotic behavior, and can be extended for two  and three-dimensional systems. We also can explore the noise effects in the quantum trajectories after adding an external random force instead of a deterministic one.

\section*{Acknowledgments}
We acknowledge fruitful remarks by W. B. de Lima and P.  de Fabritiis,  as well as partial financial support from Conselho Nacional de Desenvolvimento Científico e Tecnológico (CNPq).
\section*{Author contributions}

All the authors are responsible for the concept, design, execution, and physical interpretation of the research. 

 \section*{Declaration of Competing interests}
The authors declare no competing interests.

\end{document}